\documentclass[longauth]{aa} 

\usepackage{amsmath}
\usepackage{subfigure}
\usepackage{graphicx}
\usepackage{nicefrac}
\usepackage{xcolor}
\usepackage{lipsum}
\usepackage{caption}
\usepackage{dblfloatfix}    
\usepackage{txfonts}
\usepackage[hidelinks]{hyperref}

\usepackage{soul}

\newcommand{\hmpc}{h^{-1}\mathrm{Mpc}}

\begin{document}

   \title{Emulating galaxy and peculiar velocity clustering \\ on non-linear scales}
\author{
T.~Dumerchat \inst{1,2}\thanks{E-mail: \href{mailto:tdumerchat@ifae.es}{tdumerchat@ifae.es}},
J.~Bautista \inst{1},
C.~Ravoux \inst{3},
J.~Aguilar \inst{4},
S.~Ahlen \inst{5},
S.~BenZvi \inst{6},
D.~Bianchi \inst{7,8},
D.~Brooks \inst{9},
T.~Claybaugh \inst{4},
A.~de la Macorra \inst{10},
P.~Doel \inst{9},
S.~Ferraro \inst{4,11},
J.~E.~Forero-Romero \inst{12,13},
E.~Gaztañaga \inst{14,15,16},
S.~Gontcho A Gontcho \inst{17},
G.~Gutierrez \inst{18},
C.~Hahn \inst{19},
C.~Howlett \inst{20},
M.~Ishak \inst{21},
R.~Joyce \inst{22},
D.~Kirkby \inst{23},
A.~Kremin \inst{4},
C.~Lamman \inst{24},
M.~Landriau \inst{4},
L.~Le~Guillou \inst{25},
M.~Manera \inst{26,2},
R.~Miquel \inst{27,2},
S.~Nadathur \inst{15},
W.~J.~Percival \inst{28,29,30},
F.~Prada \inst{31},
I.~P\'erez-R\`afols \inst{32},
G.~Rossi \inst{33},
E.~Sanchez \inst{34},
D.~Schlegel \inst{4},
M.~Schubnell \inst{35,36},
J.~Silber \inst{4},
D.~Sprayberry \inst{22},
G.~Tarl\'{e} \inst{36},
B.~A.~Weaver \inst{22},
and H.~Zou \inst{37}
\begin{center} (DESI Collaboration) \end{center}
}

\institute{\scriptsize
Aix Marseille Univ, CNRS/IN2P3, CPPM, Marseille, France 
\and Institut de F\'{i}sica d’Altes Energies (IFAE), The Barcelona Institute of Science and Technology, Edifici Cn, Campus UAB, 08193, Bellaterra (Barcelona), Spain 
\and Universit\'{e} Clermont-Auvergne, CNRS, LPCA, 63000 Clermont-Ferrand, France 
\and Lawrence Berkeley National Laboratory, 1 Cyclotron Road, Berkeley, CA 94720, USA 
\and Department of Physics, Boston University, 590 Commonwealth Avenue, Boston, MA 02215 USA 
\and Department of Physics \& Astronomy, University of Rochester, 206 Bausch and Lomb Hall, P.O. Box 270171, Rochester, NY 14627-0171, USA 
\and Dipartimento di Fisica ``Aldo Pontremoli'', Universit\`a degli Studi di Milano, Via Celoria 16, I-20133 Milano, Italy 
\and INAF-Osservatorio Astronomico di Brera, Via Brera 28, 20122 Milano, Italy 
\and Department of Physics \& Astronomy, University College London, Gower Street, London, WC1E 6BT, UK 
\and Instituto de F\'{\i}sica, Universidad Nacional Aut\'{o}noma de M\'{e}xico,  Circuito de la Investigaci\'{o}n Cient\'{\i}fica, Ciudad Universitaria, Cd. de M\'{e}xico  C.~P.~04510,  M\'{e}xico 
\and University of California, Berkeley, 110 Sproul Hall \#5800 Berkeley, CA 94720, USA 
\and Departamento de F\'isica, Universidad de los Andes, Cra. 1 No. 18A-10, Edificio Ip, CP 111711, Bogot\'a, Colombia 
\and Observatorio Astron\'omico, Universidad de los Andes, Cra. 1 No. 18A-10, Edificio H, CP 111711 Bogot\'a, Colombia 
\and Institut d'Estudis Espacials de Catalunya (IEEC), c/ Esteve Terradas 1, Edifici RDIT, Campus PMT-UPC, 08860 Castelldefels, Spain 
\and Institute of Cosmology and Gravitation, University of Portsmouth, Dennis Sciama Building, Portsmouth, PO1 3FX, UK 
\and Institute of Space Sciences, ICE-CSIC, Campus UAB, Carrer de Can Magrans s/n, 08913 Bellaterra, Barcelona, Spain 
\and University of Virginia, Department of Astronomy, Charlottesville, VA 22904, USA 
\and Fermi National Accelerator Laboratory, PO Box 500, Batavia, IL 60510, USA 
\and The University of Texas at Austin, Department of Astronomy
\and School of Mathematics and Physics, University of Queensland, Brisbane, QLD 4072, Australia 
\and Department of Physics, The University of Texas at Dallas, 800 W. Campbell Rd., Richardson, TX 75080, USA 
\and NSF NOIRLab, 950 N. Cherry Ave., Tucson, AZ 85719, USA 
\and Department of Physics and Astronomy, University of California, Irvine, 92697, USA 
\and The Ohio State University, Columbus, 43210 OH, USA 
\and Sorbonne Universit\'{e}, CNRS/IN2P3, Laboratoire de Physique Nucl\'{e}aire et de Hautes Energies (LPNHE), FR-75005 Paris, France 
\and Departament de F\'{i}sica, Serra H\'{u}nter, Universitat Aut\`{o}noma de Barcelona, 08193 Bellaterra (Barcelona), Spain 
\and Instituci\'{o} Catalana de Recerca i Estudis Avan\c{c}ats, Passeig de Llu\'{\i}s Companys, 23, 08010 Barcelona, Spain 
\and Department of Physics and Astronomy, University of Waterloo, 200 University Ave W, Waterloo, ON N2L 3G1, Canada 
\and Perimeter Institute for Theoretical Physics, 31 Caroline St. North, Waterloo, ON N2L 2Y5, Canada 
\and Waterloo Centre for Astrophysics, University of Waterloo, 200 University Ave W, Waterloo, ON N2L 3G1, Canada 
\and Instituto de Astrof\'{i}sica de Andaluc\'{i}a (CSIC), Glorieta de la Astronom\'{i}a, s/n, E-18008 Granada, Spain 
\and Departament de F\'isica, EEBE, Universitat Polit\`ecnica de Catalunya, c/Eduard Maristany 10, 08930 Barcelona, Spain 
\and Department of Physics and Astronomy, Sejong University, 209 Neungdong-ro, Gwangjin-gu, Seoul 05006, Republic of Korea 
\and CIEMAT, Avenida Complutense 40, E-28040 Madrid, Spain 
\and Department of Physics, University of Michigan, 450 Church Street, Ann Arbor, MI 48109, USA 
\and University of Michigan, 500 S. State Street, Ann Arbor, MI 48109, USA 
\and National Astronomical Observatories, Chinese Academy of Sciences, A20 Datun Road, Chaoyang District, Beijing, 100101, P.~R.~China 
}

\abstract{
    We explore the potential of cross-correlating galaxies and peculiar velocities on non-linear scales to enhance cosmological constraints. 
    Leveraging the \textsc{AbacusSummit} simulation suite and the halo occupation distribution (HOD) formalism, we train emulator models to describe the non-linear clustering of galaxies and velocities in redshift space.
    Our analysis demonstrates that combining galaxy and peculiar velocity clustering, provides tighter constraints on both HOD and cosmological parameters, particularly on $\sigma_8$ and $w_0$.
    We further apply our models to realistic mock catalogues, reproducing the 
    expected density and peculiar velocity errors of type-Ia supernovae and 
    Tully-Fisher/fundamental plane measurements for the combined ZTF and DESI measurements.
    While systematic biases arise in the HOD parameters, the cosmological constraints remain unbiased, yielding $3.8\%$ precision measurement on $f\sigma_8$ compared to $4.7\%$ using galaxy clustering alone. 
    We demonstrate that, while combining tracers with realistic velocity measurements still yields improvement, the gains are diminished, highlighting the need for further efforts to reduce velocity measurement uncertainties and correct observational systematics on small scales.
}

\keywords{Cosmology -- Large scale structures -- Peculiar velocity -- Machine learning -- Multi-probe
           }
\authorrunning{T. Dumerchat et al. (DESI Collaboration)}
\titlerunning{Emulating galaxy and peculiar velocity clustering on non-linear scales}
\maketitle
%
\section{Introduction}

The large-scale distribution of galaxies in the Universe encodes valuable cosmological information, allowing us to constrain fundamental parameters governing cosmic growth and evolution. 
This is typically achieved through the analysis of clustering statistics, with the two-point correlation function (or equivalently power spectrum in Fourier space) being the most widely used observable.

Ongoing galaxy surveys such as Euclid 
(\cite{laureijs_euclid_2011}) and 
the Dark Energy Spectroscopic Instrument (DESI, \cite{desicollaborationOverviewInstrumentationDark2022})
will detect tens of millions of galaxies, significantly reducing statistical uncertainties in the galaxy two-point functions.
Yet, because the distribution of galaxies traces the non-linear and non-Gaussian matter field
(enhanced at small scales and low redshift), the two-point statistics fails to capture the full available information from the galaxy density field.
The constraining power will no longer be limited by statistics but rather by the use of those standard observables and the systematic uncertainties associated to the modelling of non-linear clustering.
Addressing this challenge requires modelling the non-linear clustering on smaller scales, or 
going beyond two point statistics.

The first point can be achieved through emulation of cosmological simulations.
In recent years, several suites of N-body simulations have been produced for this purpose (\cite{derose_aemulus_2019,villaescusa-navarro_quijote_2020, maksimova_abacussummit_2021}).
Several emulators have been developed (\cite{kobayashiAccurateEmulatorRedshiftspace2020}) and applied to real data to extract cosmological parameters (\cite{kobayashiFullshapeCosmologyAnalysis2022, chapmanCompletedSDSSIVExtended2022}).

Furthermore, alternative methods introducing summary statistics beyond the standard two-point function have been studied.
Recently, different alternative methods based on various statistics such as bispectrum analysis (\cite{Yankelevich_bispectrum_2019,Agarwal_bispectrum_2021}),
wavelet scattering transform (\cite{Valogiannis_2022}), 
and k-th nearest neighbour statistics (\cite{Yuan_knn_2023}), have been developed to extract information beyond the standard two-point function.
A promising approach, known as field-level inference (\cite{stadler_fieldlevel_2023}), consists in directly fitting 
the galaxy density field without using any compressed statistics 
using a parametrised set of initial conditions.

At low redshift
cosmic variance is the primary source of uncertainty in measurements of galaxy two-point functions.
Additional information can then be extracted from the standard two-point statistics by cross correlating different probes of the matter field
 in an overlapping volume such as cosmic voids, weak lensing, 
galaxy multi-tracer analysis (\cite{wang_MultitracerReview_2020}), density splits (\cite{paillas_2023}),
or direct measurements of the galaxy peculiar velocities.
Recent studies have shown that combining galaxy and peculiar velocity clustering 
increases the precision on growth rate measurements \citep{adamsImprovingConstraintsGrowth2017,turnerImprovingEstimatesGrowth2021,turnerLocalMeasurementGrowth2023}, leading typically to a 15-20$\%$ precision measurement of the growth rate $f\sigma_8$ when using linear modelling for scales in the range $\left[ 30,120 \right] h^{-1}$Mpc. 
The most recent measurements were recently released by the DESI Peculiar Velocity survey using Data Release 1, achieving a 12\% uncertainty by jointly analysing galaxies and peculiar velocities with three different methodologies \citep{bautistaDESIDR1Peculiar2025,carrDESIDR1Peculiar2025,douglassDESIDR1Peculiar2025,laiDESIDR1Peculiar2025,qinDESIDR1Peculiar2025,rossDESIDR1Peculiar2025,turnerDESIDR1Peculiar2025}.

In this work, we investigate the constraining power of the cross correlations of galaxies and peculiar velocities on non-linear scales.
We make use of the emulation framework introduced in \cite{Dumerchat_GP_2024} to 
jointly describe the non-linear clustering of galaxies and peculiar velocities in redshift space.
We use the \textsc{AbacusSummit} simulation suite to train our models, and build realistic mock catalogues to assess the constraining power of the combined analysis of galaxy and velocities.

This paper is organised as follows.
In section \ref{sec:data}, we present the observables to emulate, and the different data sets that we will be using.
In section \ref{sec:emu}, we describe the procedure for training the different models and validate their accuracy and precision.
In section \ref{sec:recovery}, we assess the parameters' recovery when running statistical inference.
Finally, in section \ref{sec:cosmology_constraints}, we explore the constraining power of the joint analysis on a set of mocks reproducing expected tracers density and peculiar velocity errors.


\section{Peculiar velocity and galaxy clustering from N-body simulations}
\label{sec:data} 

\subsection{The \textsc{AbacusSummit} suite of simulations}
\label{sec:data:abacus} 

The \textsc{AbacusSummit} project \citep{maksimova_abacussummit_2021} is a comprehensive set of N-body simulations performed using the Abacus N-body code \citep{garrison_abacus_2021} on the Summit supercomputer at the Oak Ridge Leadership Computing Facility. Designed specifically for the needs of the Dark Energy Spectroscopic Survey (DESI), these simulations cover an interesting range of cosmological models. The suite consists of hundreds of cubic boxes, simulated in comoving coordinates, and evolved from a high redshift of $z=8.0$ down to $z=0.1$.
Each base simulation contains $6912^3$ particles, with individual masses of $2\times 10^9 ~ h^{-1} M_\odot$, within a cubic volume of $(2~ h^{-1}\mathrm{Gpc})^3$. To identify halos, the simulations employ the \textsc{CompaSO} halo finder \citep{hadzhiyska_textsccompaso_2021}, which groups particles based on a spherical overdensity criterion. The halo catalogues generated from these simulations serve as key data products for various analyses.

\textsc{AbacusSummit} explores cosmologies around the best-fit $\Lambda$CDM model derived from Planck 2018 data \citep{planck_collaboration_planck_2020}. A total of 9 cosmological parameters are varied: $\omega_\mathrm{cdm}$ (dark matter density), $\omega_\mathrm{b}$ (baryon density), $\sigma_8$ (clustering amplitude), $w_0$ and $w_a$ (dark energy equation of state parameters), $h$ (Hubble parameter), $n_\mathrm{s}$ (spectral tilt), $N_\mathrm{ur}$ (number of massless relics), and $\alpha_\mathrm{s}$ (running of the spectral index). All cosmologies assume zero curvature.
In most cases, $h$ is chosen to match the CMB acoustic scale and could, in principle, be inferred from the other parameters. We nevertheless treat $h$ as an independent parameter, noting that the resulting constraints on this parameter should largely be prior-driven rather than coming from genuine sensitivity of the considered summary statistics.

In order to train our emulator model, we separate the available simulations into a train and a test set.
The training set covers a large parameter space, consisting of 88 distinct cosmologies. For simplicity, we denote the training set's cosmological parameters as $X_{\Omega}$, forming an $88 \times 9$ matrix.
The test simulations is composed of 6 different cosmologies, including a reference cosmology based on the Planck2018 $\Lambda$CDM model with 60 meV neutrinos, alongside variations such as a model with massless neutrinos, lower $\omega_\mathrm{cdm}$ based on WMAP7 data, a $w$CDM model, and variations in $N_\mathrm{ur}$ and $\sigma_8$.

All simulations share identical initial conditions, ensuring that variations in clustering measurements arise solely from changes in cosmology. To evaluate cosmic variance, \textsc{AbacusSummit} also contains a set of 25 realisations for a single cosmology but with varied initial conditions.
To estimate covariance matrices, \textsc{AbacusSummit} provides a set of 1400 smaller boxes, each of volume $\rm(500~\hmpc)^3$, with matching cosmology and mass resolution but different initial conditions.
All these simulations were ran using the baseline Planck2018 cosmology. Although snapshots are available for multiple redshifts, this work exclusively uses snapshots at $z=0.2$ as it is the lowest redshift with corresponding small box realisations. We do not expect our results to depend on this choice.

\subsection{Halo occupation distribution model}
\label{sec:data:hod} 

To populate dark-matter halos in our simulations with galaxies, we adopt the halo occupation distribution (HOD) model \citep{zheng_galaxy_2007}. This model defines the number of galaxies $N$ within a given halo as a stochastic variable, following a probability distribution $P(N|\Omega_h)$ where $\Omega_h$ is a set of halo characteristics. For this work, we use the standard HOD model, where the galaxy number distribution is solely dependent on halo mass ($\Omega_h = M$). 
The galaxy occupation is split into two components: central and satellite galaxies, with the total occupation given by $\langle N(M) \rangle = \langle N_{\rm cen}(M) \rangle + \langle N_{\rm sat}(M) \rangle$. The number of central galaxies is drawn from a Bernoulli distribution, while the satellite galaxy count follows a Poisson distribution. The means of these distributions are defined as:

\begin{equation}
\begin{aligned}
    \langle N_{\rm cen} \rangle &= \frac{1}{2} {\rm erfc}\left( \frac{{\rm log_{10}}(M_{\rm cut}/M)}{\sqrt{2}\sigma}\right),\\
\langle N_{\rm sat} \rangle &= \left [ \frac{M - \kappa M_{\rm cut}}{M_1} \right]^{\alpha}\langle N_{\rm cen} \rangle,
\end{aligned}
\label{eq:hod_cen_sat}
\end{equation}

Here, erfc($x$) represents the complementary error function, and the HOD parameters $\theta_{\rm hod} = \{M_{\rm cut}, \sigma, \kappa, M_1, \alpha \}$ are defined as follows: $M_{\rm cut}$ is the minimum halo mass for hosting a central galaxy, $\sigma$ represents the width of the cutoff profile (or the sharpness of the transition from 0 to 1 central galaxies), $\kappa M_{\rm cut}$ is the minimum halo mass for hosting satellites, $M_1$ is the characteristic mass for hosting one satellite, and $\alpha$ is the power-law index governing satellite occupation.

\begin{table}
\caption{Cosmological and HOD parameter ranges sampled for building the emulator's training and testing datasets. Cosmological bounds are based on the smallest hypercube around the AbacusSummit grid, while HOD bounds reflect the minimum and maximum values found across 600 HOD configurations.}
\centering
\begin{tabular}{ccc}
\hline
\hline
  & 
  Parameter &
  Range \\
  \hline
  
Cosmology & $\rm \omega_{cdm}$& [0.103 ,0.140] \\
 & $\rm \omega_{b}$& [0.0207,0.0243 ] \\
& $\sigma_8$& [ 0.678,0.938] \\
& $\rm w_0$& [-1.271,-0.726]\\
& $\rm w_a$& [-0.628,0.621] \\
& $\rm h$& [  0.575 , 0.746 ] \\
& $\rm n_s$& [ 0.901, 1.025] \\
& $\rm N_{ur}$& [1.020 , 3.046 ]\\
& $\rm \alpha_s$& [-0.038 , 0.038] \\

\hline

HOD &$\rm \alpha$& [0.30, 1.48] \\
& $\rm \kappa$&  [0.00, 0.99]\\
 & $\rm log_{10}M_1$&  [13.6, 15.1]\\
 & $\rm log_{10}M_{cut}$& [12.5, 13.7]  \\
 & $\rm log_{10}\sigma$& [-1, 0] \\

\hline
\hline
\end{tabular}
\label{tab:cosmo_hod_range}
\end{table}

The galaxy assignment is performed using the \textsc{AbacusHOD} implementation \citep{yuan_abacushod_2022} where, unlike traditional methods that use a Navarro-Frenk-White profile \citep{navarro_universal_1997}, satellite galaxies are assigned to particles of the halo (with the corresponding velocity) based on the distribution of dark matter particles in the halo.
Central galaxies are assigned to the centre of mass of the halo (with the corresponding velocity).

This halo occupation framework is publicly available in the \textsc{abacusutils} \footnote{\url{https://github.com/abacusorg/abacusutils}} package.

For the training set, we populate the 88 simulation boxes with 600 HOD models, selected via Latin hypercube sampling. 
In contrast, the test set comprises 20 different HOD models applied to the 6 test cosmologies. 
The parameter ranges for these HODs are consistent with measurements of the Bright Galaxy Sample from the DESI one percent survey \citep{prada_desi_2023}. In the rest of this paper, we refer to $X_{\rm HOD}$ as the $600 \times 5$ matrix containing the HOD parameters for the training set.

To reduce shot noise, we repeat the galaxy assignment process five times for the same HOD model, only changing the random seed. The clustering results are then averaged. Since the same HOD configurations are used across all cosmologies, this creates a parameter space structured as $X_{\Omega} \otimes X_{\rm HOD}$, where $\otimes$ is the Kronecker product. This particular structure will be essential for our emulator construction discussed in section \ref{sec:emu:mkgp}.

\subsection{Observable}
\label{sec:data:observable} 

The observables we which to emulate on the non-linear scales are the galaxy-galaxy, velocity-velocity and velocity-galaxy two point correlation functions (TPCF) in redshift space.
Based on a pair-counting strategy, the Landy-Szalay estimator \citep{landy-szalay} is commonly used to describe the galaxy auto-correlation function 
\begin{equation}
\xi_{gg}(\textbf{r}) = \frac{DD(\textbf{r}) -DS(\textbf{r}) - SD(\textbf{r}) }{SS(\textbf{r})} + 1,
\label{eq:LS_gg}
\end{equation}
where $\textbf{r}$ is the pair separation vector, $DD$, $SS$, $DS$ and $SD$ are the normalised pair counts of galaxy-galaxy, random-random, galaxy-random and random-galaxy catalogues. The random sample catalogue is used to define the window function of the survey. 
We derive analogous estimators for the galaxy weighted peculiar velocity auto and cross-correlation functions
\begin{equation}
\xi_{vv}(\textbf{r}) = \frac{VV(\textbf{r})}{RR(\textbf{r})} ,\\
\xi_{vg}(\textbf{r}) = \frac{VD(\textbf{r}) - VS(\textbf{r})}{RS(\textbf{r})} ,
\label{eq:LS_vv_vg}
\end{equation}
with $VV$, $RR$, $VS$ and $RS$ the normalised pair counts of the peculiar velocities, galaxies and random catalogues.
Note that in this definition there are two random catalogues $S$ and $R$, one for the galaxy sample and the other for the peculiar velocity sample, which are not necessarily the same. In this work however, we use the same volumes and homogeneous distributions for both samples. Thus, accounting for the overall number count normalisation we can use just one random catalogue $S = R$.
We provide a complete derivation of our estimators in appendix \ref{sec:annex:estimator}. 
Before measuring the clustering, we move every galaxy to its redshift-space position by applying the shift along the line of sight (LoS)
\begin{equation}
d_s = d + \frac{v}{H(z)}(1+z),
\label{eq:rsd}
\end{equation}
with $d$ and $d_s$ the true and distorted radial position, $v$ the peculiar velocity along the LoS, and $H(z)$ the Hubble parameter at the considered redshift. 
As we are working with snapshots, the redshift evolution is not simulated and $z$ has the same value of 0.2 for every galaxy.
Once measured, the TPCF are decomposed as a series of Legendre polynomials :
\begin{equation}
\xi(s,\mu) = \sum_l P_l(\mu) \xi_l(s) ,
\label{eq:legendre1}
\end{equation}
where
\begin{equation}
\xi_l(s) = \frac{2l+1}{2}\int \xi(s,\mu) P_l(\mu) d\mu ,
\label{eq:legendre2}
\end{equation}
with $s$ the separation between the galaxy pairs and
$\mu$ the cosine of the angle between the observer LoS and the separation vector of the galaxy pair.
In the following, we will use the monopole and quadrupole for $\xi_{gg}$ and $\xi_{vv}$ and the dipole for $\xi_{vg}$. Note that while $\xi_{vg}$ is neither an odd nor even function of $\mu$, we find that the dipole gives the stronger signal.
We compute those observables multiple times for several cosmologies and HODs in order to build our emulator training set.
The most expensive part is the random-random pair counting. 
Because we are working with cubic boxes with periodic boundary conditions, under the flat-sky approximation, $SS$ can be analytically computed and we can make use of the natural estimator \citep{peebles} and its analogue for peculiar velocities :
\begin{equation}
\begin{aligned}
&\xi_{gg}(\textbf{r}) = \frac{DD(\textbf{r})}{SS(\textbf{r})} - 1, \quad \xi_{vv}(\textbf{r}) = \frac{VV(\textbf{r})}{SS(\textbf{r})} ,\\  \\
&\xi_{vg}(\textbf{r}) = \frac{VD(\textbf{r}) }{SS(\textbf{r})} ,
\end{aligned}
\label{eq:peebles}
\end{equation}
where 
\begin{equation}
SS(\textbf{r}) =\frac{N_{\text {gal }}\left(N_{\text {gal }}-1\right)}{L_{\text {box }}^3}\left[\frac{4 \pi\left(s_2^3-s_1^3\right)}{3}\right]\left[\mu_2-\mu_1\right].
\label{eq:analytic_random}
\end{equation}
With $N_{\mathrm{gal}}$ and $L_{\mathrm{box}}$ the number of galaxies and the size of the simulation box respectively, $s_1$ and $s_2$ the lower and upper limits of the separation bins, and $\mu_1$ and $\mu_2$ the lower and upper limits of the angular bins.
We verify below that this faster estimator does not yield biased measurements of the clustering, as the flat sky approximation breaks down at low redshift and large separation.

\begin{figure}
  \centering

 \includegraphics[width=0.9\columnwidth]{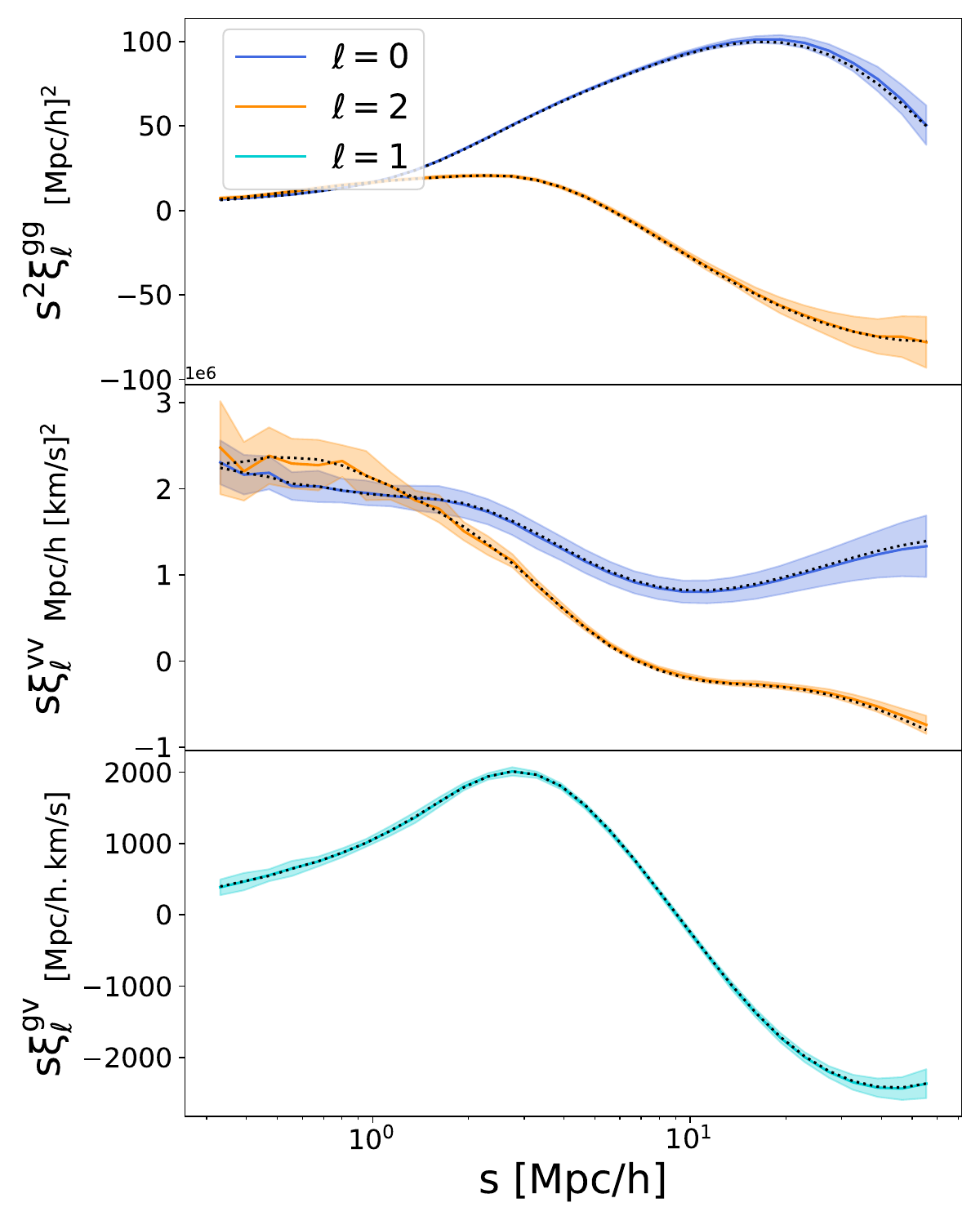}  
\caption{Comparison between fixed and varying LoS estimators using of the 25 Planck2018 boxes with the median HOD.
The solid lines show the mean clustering of the 25 boxes computed using Eqs \ref{eq:LS_gg} and \ref{eq:LS_vv_vg} with a full-sky spherical footprint and a redshift range of $z \in [0 ; 0.1] $. The shaded area are the corresponding standard deviation of the realisations. The dashed black lines show the mean of the flat-sky approximation clustering.}
\label{fig:wideangle}
\end{figure}

We measure the clustering of the 
25 Planck2018 boxes with the median HOD - defined as the centre of the sampled HOD parameters space - using a full-sky spherical footprint with a radial cut corresponding to a redshift range of $z \in [0 ; 0.1] $.
In the following we refer to those measurements 
as the recovery set.
We choose the separation range $X_\textrm{s}$ to be a logarithmic binning going from the resolution of the simulation $0.3~\hmpc$ to $60~\hmpc$.
Paircounting is performed using the package \textsc{pycorr}\footnote{\url{https://github.com/cosmodesi/pycorr}}.
In figure \ref{fig:wideangle} the solid lines show the expected mean clustering of the recovery set different multipoles (with the standard deviation in shaded area). 
The dashed black lines show the mean clustering of the 25 boxes computed using Eq \ref{eq:peebles} (fixed LoS).
We see that for the scales considered, the deviation from the expected value of the flat-sky approximation measurements are within the cosmic variance, so we can safely neglect wide angle effects in the following.

\begin{figure}
  \centering
 \includegraphics[width=1.\columnwidth]{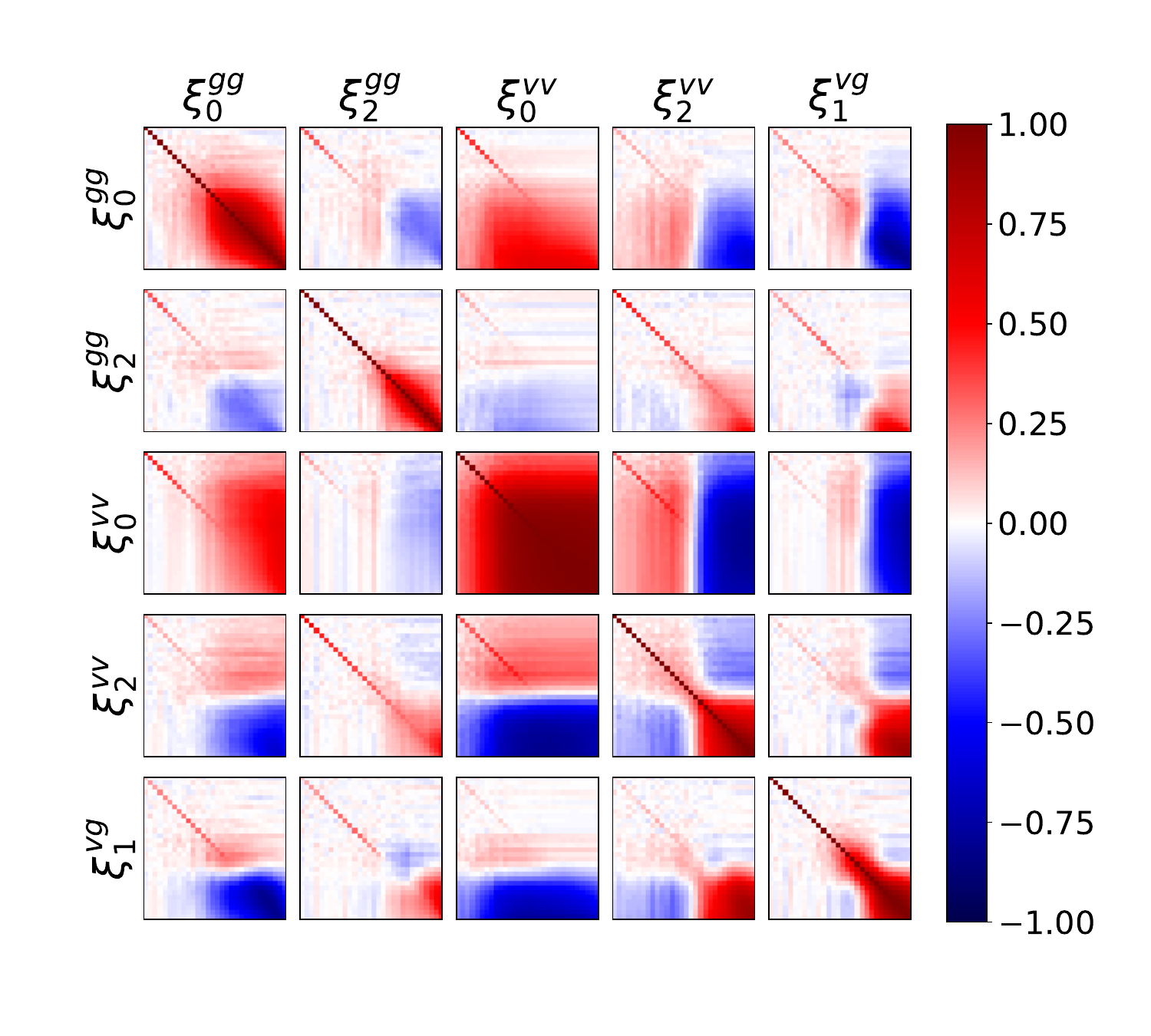}  
\caption{Correlation matrix computed from 1400 small boxes realisations of the Planck2018 cosmology, using the median HOD. The different panels show the auto and cross correlation between the TPCF multipoles of $\xi_{gg}$ ,  $\xi_{vv}$ and  $\xi_{gv}$. Each of the subplot axis correspond to the scale $s$ with a logarithmic binning going from $0.3~\hmpc$ to $60~\hmpc$.}
\label{fig:corr_matrix}
\end{figure}

We use Eq \ref{eq:peebles} to measure our training and testing sets.
While the flat-sky approximation enable to use the full volume of the cubic boxes and reduce the effect of cosmic variance, we still need to account for it in our measurements since the train and test cosmology boxes are run with the same initial conditions.
Using the same estimators, we build the covariance matrix $C_{{\rm cosmic}}$ from the 1400 small boxes for Planck2018 cosmology with the median HOD. We then rescale the covariance to match the volume of the baseline boxes.
Figure \ref{fig:corr_matrix} gives the corresponding correlation matrix between the different observables for our selected separations bins.
We find strong correlations for the velocity statistics, over a wide range of scales, especially for the velocity auto-correlation monopole leading to smooth variation of the signal.

\subsection{Realistic peculiar velocity catalogue}
\label{sec:data:realistic}

In every TPCF measurement considered so far, every galaxy position was associated with a radial peculiar velocity.
Yet it is well known that measuring peculiar velocities is a tricky task requiring both 
a redshift and a redshift-independent distance indicator.
Hence peculiar velocity surveys are typically less complete than galaxy surveys.
The main techniques to measure peculiar velocity in large scale spectroscopic survey rely on using the Tully-Fisher(TF) and Fundamental Plane (FP) \citep{campbell_6df_2014,hong_2mtf_2019,howlett_sloan_2022,saulder_target_2023}.
Those are empirical relationships between photometric and spectroscopic quantities of spiral and elliptical galaxies respectively, allowing to derive an estimate of the distance.
The upcoming Rubin Observatory and the Zwicky Transient Facility (ZTF) \citep{ZTF_2019}
will produce samples of type Ia Supernovae (SnIa) and provide additional distance estimates, 
in smaller number but with greater precision. 
We want to create a set of 'realistic' mocks to study the 
impact of distance uncertainties and number densities of peculiar velocity tracers on the cosmological constraints. 

We use the 25 boxes with Planck2018 cosmology to create three distinct
catalogues: a galaxy sample tracing the density field only, a joint TF and FP velocity sample (similar uncertainties), 
and a separate SnIa velocity sample. 
We consider the SnIa and TF/FP samples as a single velocity catalogue hereafter.
We choose an HOD that roughly matches the expected number densities $n(z)$ within $z \in [0,0.1]$.
We get a galaxy density around $\Bar{n}(z) = 2 \times 10^{-3} h^{-3}/\text{Mpc}^3$  (correspond to a subsample of the DESI Bright Galaxy Survey). 
From this sample, we randomly select galaxies to get a TF/FP sample with a density around $\Bar{n}(z) = 10^{-3} h^{-3}/\text{Mpc}^3$,
 and a SnIa sample with a density around $\Bar{n}(z) = 10^{-4} h^{-3}/\text{Mpc}^3$.
The corresponding HOD parameters used to populate the galaxy sample are 
$ \alpha = 1.3$,
$ \kappa = 0.8$,
$ \log_{10}M_1 = 13.7$,
$ \log_{10}M_\textrm{cut} = 12.6$ and 
$ \log_{10}\sigma = -0.35$.
Furthermore, in order to test the accuracy of our emulation framework under Alcock-Paczynski (AP) distortions (\cite{alcock-paczynski}), we chose a fiducial cosmology(\textsc{c003} in \textsc{AbacusSummit}) different from the true one to convert redshifts into comoving distances.

Peculiar velocity surveys typically measure the log distance ratio defined as $\eta \equiv \log_{10}(D(z_{obs})/D_{obs} ) $, where $D(z_{obs})$ is the comoving radial distance calculated for the observed redshift (this term contains the peculiar velocity contribution), and $D_{obs}$ is the distance estimated using SnIa or TF/FP methods, which is an indicator of the true cosmological distance (without the influence of the peculiar velocity). 
The log distance ratio can be expressed as a function of peculiar velocity $v$ (see Appendix \ref{annex:vel})
\begin{equation}
    \eta = \alpha(z_\textrm{ obs}) v ,
\end{equation}
with the proportionality factor defined as 
\begin{equation}
    \alpha(z_\textrm{ obs}) \equiv \frac{1}{\ln 10}  \frac{(1+ z_\textrm{ obs})}{D(z_\textrm{ obs})H(z_\textrm{ obs})}.
\label{eq:alpha_pv_factor}
\end{equation}
Because of the intrinsic scatter of standardisable objects, the log-distance ratio measurements have a typical error 
$\sigma_\eta = \left (f_\textrm{ err} / \ln10 \right )$ with $f_\textrm{ err}$ around 20 \% for TF/FP, and 7\% for SnIa (\cite{boruah_cosmic_2020}).
After applying the redshift cut, we convert the peculiar velocities to log distance ratio. 
A Gaussian random error is then applied on $\eta$ with zero mean and a standard deviation corresponding to a $f_\textrm{ err}$ of 7\% and 20 \% for the SnIa and TF/FP samples respectively.
Note that this uncorrelated noise is an approximation.
In real analyses, the distances errors might be correlated due to calibration or environmental dependencies.
As $\alpha(z)$ is an decreasing function of the redshift, the peculiar velocity errors $\sigma_v$ increase with redhsift.

While FKP weighting \citep{feldman_power_1994} is commonly used for the galaxy density sample, similar weights can be used to avoid biasing the velocity clustering measurements \citep{qin_redshift-space_2019}.
We weight the galaxy and velocity samples using
\begin{equation}
w_{g} = \frac{1}{P_{g} \ \bar n(z)  + 1} \quad \text{and} \quad  w_{v} = \frac{1}{P_{v} \ \bar n(z)  + \left(\sigma_\eta / \alpha(z)\right)^2} ,
\label{eq:weights_fkp}
\end{equation}
setting $P_g=10^4 \ h^{-3}  \textrm{Mpc}^3 $ and $ P_v = 10^9 \ h^{—3}  \textrm{Mpc}^3 \ \textrm{km}^2  \textrm{s}^{-2}$.

%
%
%

\begin{figure}
  \centering
 \includegraphics[width=0.9\columnwidth]{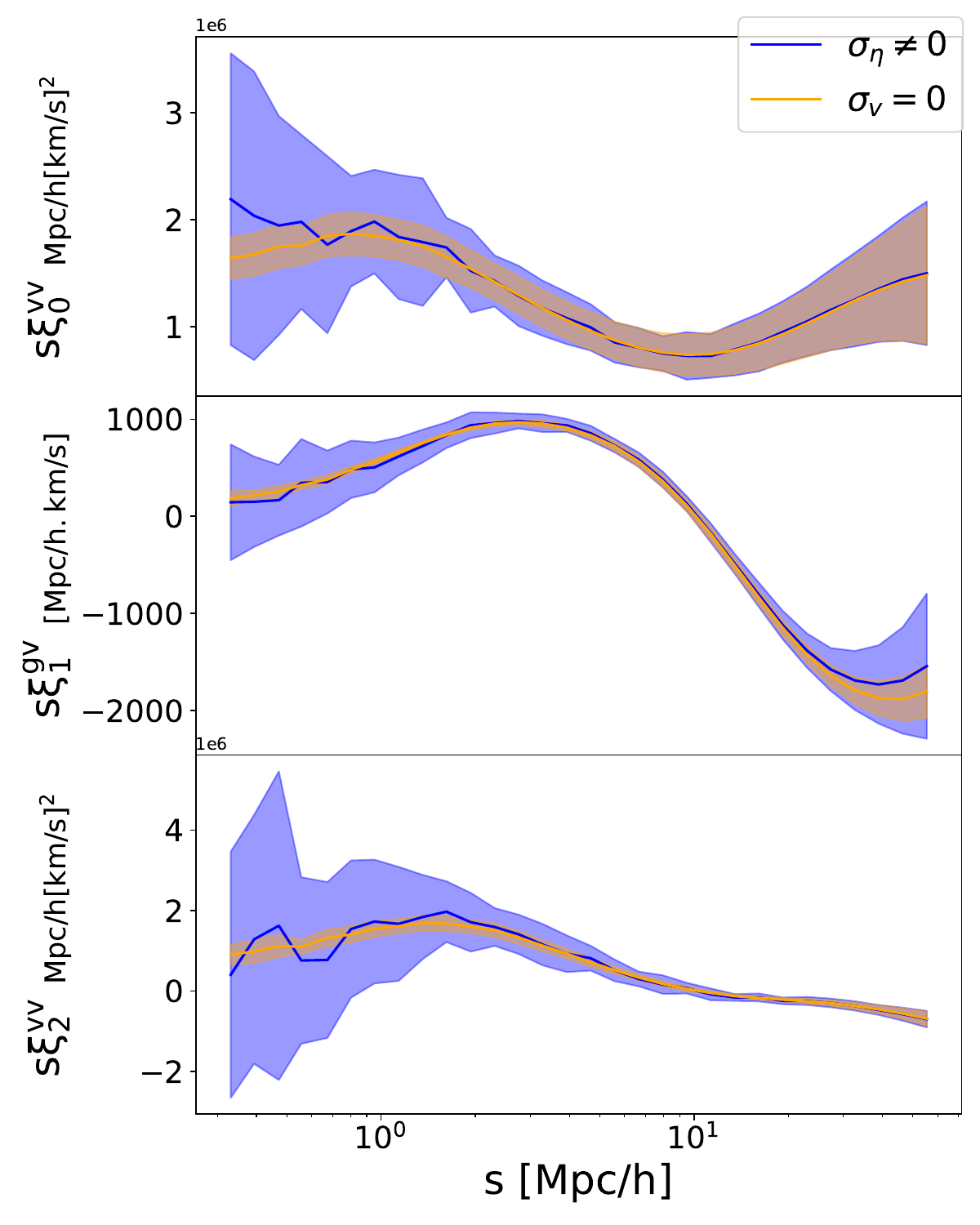}  
\caption{Comparison between the noisy and the noise-free measurements for peculiar velocity clustering using our set of 25 realistic mocks.
The average of the 25 "true" clustering is shown in orange and our estimator is shown in blue. The shaded area is the standard deviation of the 25 mocks.}
\label{fig:pv_noise_estimator}
\end{figure}

Figure \ref{fig:pv_noise_estimator} compares the average of the 25 ``true'' clustering (in orange) with the noisy, weighted, measurements (in blue). The shaded areas correspond to the standard deviation over the 25 realisations.
Our noisy clustering measurements are unbiased with respect to the noise-free measurements.
As expected, the lower density of tracers increases the uncertainties on small scales (larger shot-noise).
$\xi^{vv}$ multipoles variance remains unaffected for the largest scales.
We further
build a covariance matrix for our realistic mocks applying the same procedure to the 1400 small
boxes.

We summarise the different clustering measurements that will be used in the following sections:
\begin{itemize}
\item[$\bullet$]  Training set : $N_\text{cosmo} \times N_\text{HOD} =  88 \times 600$ different clustering with fixed ICs. Used to train the emulators.
\item[$\bullet$]  Test set : $N_\text{cosmo} \times N_\text{HOD} =  6 \times 20$ different clustering with fixed ICs. Used to assess the emulators' performance.
\item[$\bullet$]  Recovery set : $25$ realisations of the Planck2018 cosmology with the median HOD and a spherical full sky footprint for a redshift range $z \in [0,0.1]$. 
Used to run MCMC and validate the recovery on cosmological and HOD parameters using the different emulators. 
\item[$\bullet$] Realistic mocks : $25$ realisations of the Planck2018 cosmology with Gaussian log-distance ratio errors and a HOD matching the expected $n(z)$.
Full sky footprint for a redshift range $z \in [0,0.1]$ and AP effect. Used to test the constraining power of the emulator on realistic setting.
\item[$\bullet$] Small Boxes : 1400 realisations of the Planck2018 cosmology. Used to build covariance matrices for the training set and the realistic mocks.
\end{itemize}

\section{Building the emulators}
\label{sec:emu}

The emulator training procedure aims at building
the mapping function between the input $X_\Omega$, $X_\textrm{ HOD}$, $X_s$ and our output: the multipoles $\xi^{ab}_\ell$ using our training set.
An important caveat here is that the mapping function 
depends on the specificities of our simulation suite.
Therefore, one would require running different types of n-body simulations to properly estimate the confidence level of our model.
Additionally, it is important to estimate the precision of the model when interpolating away from the training points.
Gaussian Process formalism allows to estimate a mean prediction for some new input values and the corresponding covariance between the newly predicted points, while considering the uncertainties in the training set.
In the following we describe the emulation procedure and discuss the performance of the trained models for the different observables.


\subsection{Multi-scale Gaussian process}
\label{sec:emu:mkgp} 

We build independent emulators for our five statistics $\left [ \xi^{gg}_0,\xi^{gg}_2,\xi^{vv}_0,\xi^{vv}_2,\xi^{vg}_1 \right ]$ using the multi-scale Gaussian process algorithm described in \citet{dumerchat_galaxy_2023} and publicly available as the \textsc{MKGpy}\footnote{\url{https://github.com/TyannDB/mkgpy}} library.
The grid sampling of the parameter space allows use to decompose the signal and noise kernels as Kronecker products
\begin{equation}
K = K_{\Omega} \otimes K_\textrm{HOD} \otimes K_\textrm{s}, \quad \text{and} \quad
N = N_{\Omega} \otimes N_\textrm{HOD} \otimes N_\textrm{s},
\end{equation}
where $K_{\Omega}, K_\textrm{HOD} $ and $ K_\textrm{s}$ are the kernels used to respectively model the cosmological, HOD parameters and scale dependence of the signal. The kernels $N_{\Omega} , N_\textrm{HOD} $ and $ N_\textrm{s}$ are analogously used to describe the parameter dependence of the noise in the training set.
We fix $N_\textrm{s}$ to be equal to $C_{{\rm cosmic}}$ described previously in figure \ref{fig:corr_matrix}, and for the cosmological and HOD noise kernels with chose diagonal heteroscedastic kernels allowing to rescale the variance for every HOD and cosmological configuration. 
This leads to $88 + 600 = 688$ noise hyperparameters for every multipole.
For the signal kernels, we use standard squared exponential kernels (as we found that more sophisticated kernels do not increase the emulator performance), leading to  16 signal hyperparameters composed of 9 lengthscales for the cosmological parameters, 5 for the HOD parameters, one for the separations and one overall variance of the signal.
We take the logarithm of the separation $\log_{10} X_s$ to train our emulators over a more smoothly varying signal. 
To avoid numerical overflow during training, we further normalise each input dimension $X_\Omega$, $X_{\rm HOD}$ and $\log_{10} X_s$ between 0 and 1.
To reduce the dynamical range of the output (our multipoles), we apply a bi-symmetric log transformation \citep{Webber_2013} suited for datasets with both negative and positive values, defined as 
\begin{equation}
\operatorname{slog}_S(x)=\operatorname{sgn}(x) \cdot \log _{10}(1+|(x / S)|),
\label{eq:slog}
\end{equation}
where the constant $S$ adjusts the slope near the origin.
We then normalise the output of our training sets to be centred on zero and have a standard deviation of one.
The $S$ values are chosen to fine tune the model. We set it to 10$^1$ for $\xi^{gg}_0$ and $\xi^{gg}_2$, 10$^3$ for $\xi^{vv}_0$ and $\xi^{vg}_1$, and 10$^5$ for $\xi^{gg}_2$.
The covariance $C_{{\rm cosmic}}$ used as the $N_\textrm{s}$ kernel is normalised as well following the same steps.
Once the input and output of the training set are normalised, the hyperparameters are optimised using the \textsc{Scipy} gradient based algorithm \textsc{lbfgs} with 10 random initialisations. 
When computing a new prediction, the mean and covariance are transformed back into physical space.

\subsection{Prediction accuracy and precision}
\label{sec:emu:accuracy} 

We first test our models accuracy on the test set by measuring the Median Absolute Error (MAE) for the five multipoles. 
The MAE is defined as $\operatorname{MAE} \left ( \xi\right) = \operatorname{median} \left | \left (\xi^{\rm emu} - \xi^{\rm test} \right) / {\xi^{\rm test}} \right |$.
While this metric is ill-defined when $\xi^{\rm test}$ approaches zero, it is commonly used to compare precisions between various emulators.

\begin{figure}
  \centering
 \includegraphics[width=0.9\columnwidth]{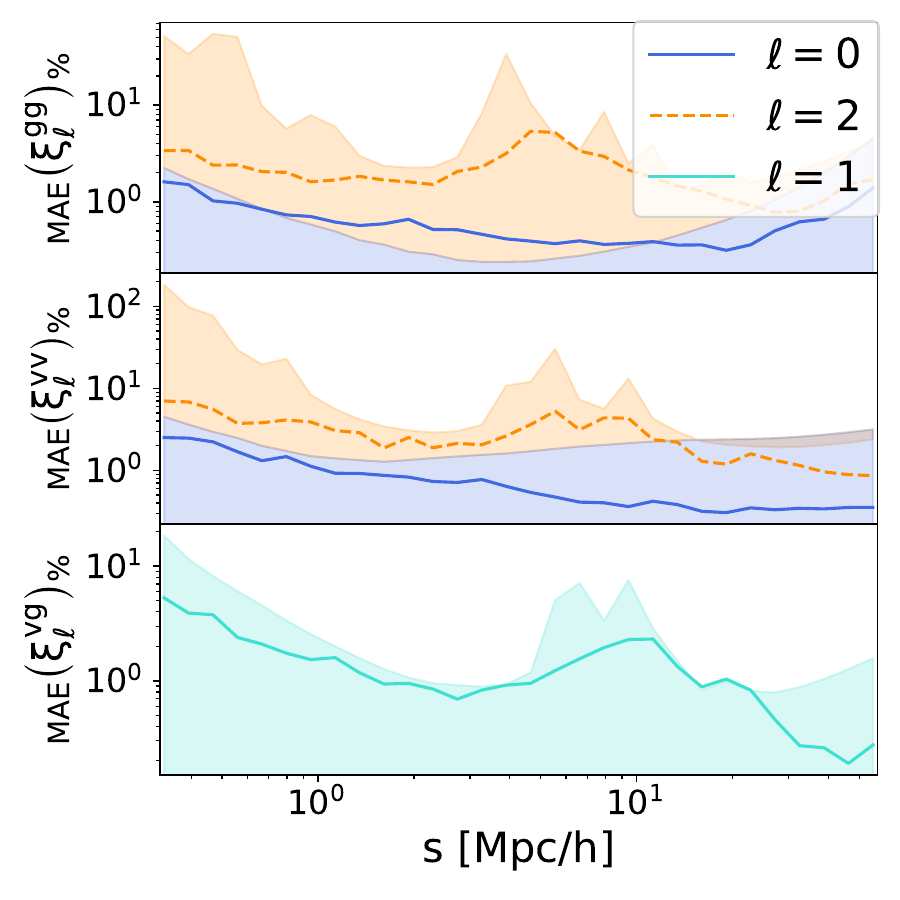}   
\caption{Median Absolute Error (MAE) evaluated of the complete test set.
The blue and orange solid lines in the top (middle) panel show the performance of the $\xi^{gg}_0$ and $\xi^{gg}_2$ ($\xi^{vv}_0$ and $\xi^{vv}_2$) models. The turquoise solid line in the bottom panel show the performance for $\xi^{vg}_1$.
The shaded area estimate the expected accuracy on the test set as the median of $| \sigma/\xi^{\rm test} |$.}
\label{fig:emu_accuracy}
\end{figure}

Figure \ref{fig:emu_accuracy} presents the MAE in percent. 
The blue solid and orange dashed lines in the top panel show the performance of the $\xi^{gg}_0$ and $\xi^{gg}_2$ models. The same colours are used for $\xi^{vv}_0$ and $\xi^{vv}_2$ in the middle panel. The turquoise solid line show the performance for $\xi^{vg}_1$ in the bottom panel.
The shaded area estimate the expected accuracy on the test set as the median of $| \sigma/\xi^{\rm test} |$ in \%, with $\sigma^2$ the diagonal of $C_{\rm cosmic}$.
While we find good performance for every models, as the MAEs roughly follow the expected accuracies, we observe some discrepancies.
For $\xi^{gg}_0$, on scales between 1-10 $\hmpc$, the MAE is slightly larger than the expected accuracy.
We also see that the models are sometime overfitting the test set, especially on $\xi^{vv}_0$ large scales as they are very much correlated and signal is smooth (see figure \ref{fig:corr_matrix}). The GP might struggle evaluating the amount of noise.
We argue that both these under and over-fitting should be taken into account by the emulator predicted covariance.
We verify this statement using the recovery set as it is no longer fixed to the same IC. For each of the 25 mocks we compute both the difference $\Delta \xi = \left (\xi^{\rm emu} - \xi^{\rm reco} \right) $ and the total covariance matrix $C_{\rm tot} = C_{\rm cosmic} + C_{\rm emu}$ with $C_{\rm cosmic}$ rescaled to match the recovery set volume.

\begin{figure}
  \centering
 \includegraphics[width=0.9\columnwidth]{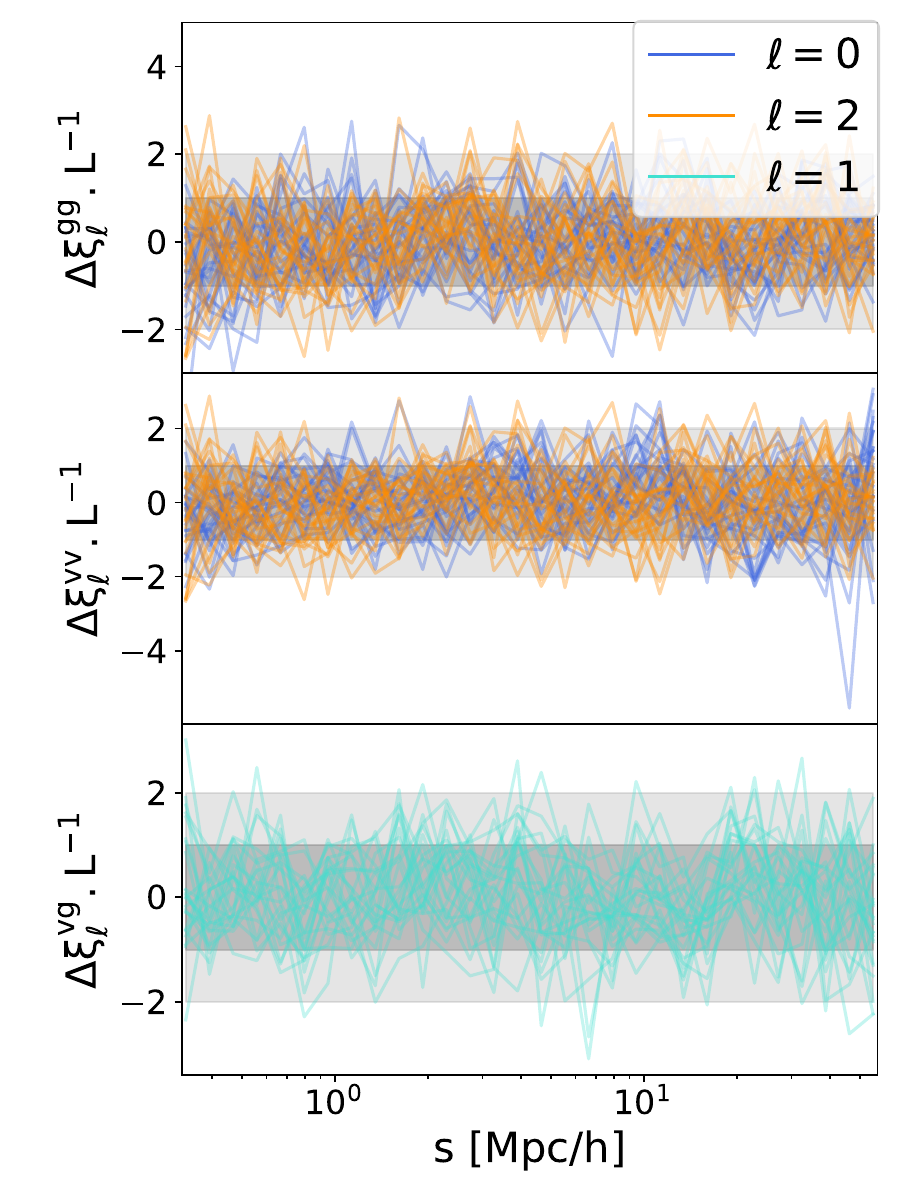}   
\caption{Residuals $\Delta \xi \cdot L^{-1} $ with $L^{-1}$ the Cholesky decomposition of the inverse of the total covariance $C_{\rm tot} = C_{\rm cosmic} + C_{\rm emu}$.
The blue and orange solid lines in the top panel show the performance of the $\xi^{gg}_0$ and $\xi^{gg}_2$ models. The same colours are used for $\xi^{vv}_0$ and $\xi^{vv}_2$ in the middle panel. The turquoise lines show the performance for $\xi^{vg}_1$ in the bottom panel.
The grey shaded areas are the one and two sigma deviations.}
\label{fig:emu_recovery}
\end{figure}

Figure \ref{fig:emu_recovery} shows the diagonalised residuals $\Delta \xi \cdot L^{-1} $ with $L^{-1}$ the Cholesky decomposition of the inverse of the total covariance (allowing to consider the correlations between scales in the residuals).
As previously, the blue and orange solid lines in the top panel show the performance of the $\xi^{gg}_0$ and $\xi^{gg}_2$ models. The same colours are used for $\xi^{vv}_0$ and $\xi^{vv}_2$ in the middle panel. The turquoise lines show the performance for $\xi^{vg}_1$ in the bottom panel.
The grey shaded areas are the one and two sigma deviations.
We find that the different models are performing reasonably well, although for $\xi^{vv}_0$ the largest scales residuals can be four sigmas large which is statistically significant out of only 25 realisations. 
However the considered covariance does not contain any observational noise and might be under estimated.
In the following test we our models by performing an inference on the cosmological and HOD parameters.

\section{Parameters recovery}
\label{sec:recovery}

In order to validate our models, we use the recovery set to perform statistical inferences over the cosmological and HOD parameters,
for different combination of observables.

\subsection{Inference strategy}
\label{sec:recovery:inference} 

We perform the parameter inference by running MCMC chains with the library \textsc{emcee} (\cite{emcee_2013}), using the flat prior corresponding to 
the train set described in table~\ref{tab:cosmo_hod_range}, and a Gaussian log-likelihood defined as
\begin{equation}
\log \mathcal{L}(\boldsymbol{\theta}) \propto -\frac{1}{2} \left(\boldsymbol{d - \mu(\theta)} \right) C^{-1}_{\textrm{ tot}}(\boldsymbol{\theta}) \left(\boldsymbol{d- \mu(\theta) } \right)^\top -\frac{1}{2} \textrm{log} |C_{\textrm{ tot}}(\boldsymbol{\theta})|
\label{eq:likelihood_sbi}
,\end{equation}
where $\boldsymbol{d}$ is the data vector, $\boldsymbol{\mu(\theta)}$ is the model prediction for a 
given set of $n_p$ parameters $\boldsymbol{\theta} = \left( \boldsymbol{\theta}_{\Omega},\boldsymbol{\theta}_\textrm{hod} \right)$, and $C_{\textrm{ tot}}(\boldsymbol{\theta})$ is the total covariance defined in section \ref{sec:emu:accuracy}.
Compared to a `standard' cosmological inference, the terms $C^{-1}_{\textrm{ tot}}(\boldsymbol{\theta})$ and $\textrm{log} |C_{\textrm{ tot}}(\boldsymbol{\theta})|$ could be computed for every iteration because of the emulator covariance dependence on the parameters space (proximity of training points). 
However, to reduce the computational cost of likelihood estimation, we fix the emulator covariance to its expected value, evaluated at the true simulation parameters.

To cross-check each observable and evaluate the amount of additional information contained in the velocity statistics,
we use different concatenation of the multipoles $\xi_\ell^{ab}$ for the data vector $\boldsymbol{d}$.
When using the full set of multipoles to perform the 
combined analysis
the data vector becomes significantly large ($n_d = 150$) with respect to the number of mocks used to estimate the covariance matrix ($n_\textrm{mock}=1400$).
Consequently, to account for the noise in the covariance and precision
matrix estimates, we rescale $C_\textrm{cosmic}$ 
with the revisited factor derived in \cite{Percival_wishart_2022}
\begin{equation}
  m_1 = \frac{(n_\textrm{mock}-1)\left[1 + B(n_d-n_p) \right]}{n_\textrm{mock} - n_d + n_p -1},
\end{equation}
with
\begin{equation}
   B = \frac{(n_\textrm{ mocks}-n_d-2)}{(n_\textrm{ mocks}-n_d-1)(n_\textrm{ mocks}-n_d-4)}.
\end{equation}

\subsection{Results}
\label{sec:recovery:results}


\begin{table*}
  \centering
  \caption{Summary statistics for the fits of the 25 spherical full sky footprint mocks with Planck2018 cosmology 
  and the median HOD. $\langle \Delta_\theta \rangle$ is the bias, the difference between the true and best fitted value for any parameter $\theta$ scaled by $10^2$.
   $\langle \sigma_\theta\rangle$ is the average over the 25 realisations of the (symmetrized) one sigma confidence level scaled by $10^2$. 
   $\sigma(Z_\theta)$ is the standard deviation of the standard score for any parameter $\theta$. 
   Note that $f\sigma_8$ is a derived parameter from the other cosmological ones, and corresponds to redshift $z=0.2$.}
   \small
  \addtolength{\tabcolsep}{-1pt}
  {
  \begin{tabular}{l|ccc|ccc|ccc|ccc}
  \hline
  \hline
     & \multicolumn{3}{c|}{gg }  & \multicolumn{3}{c|}{vv } & \multicolumn{3}{c|}{vg } & \multicolumn{3}{c}{tot }  \\
   $p$ &
  $\langle \Delta_\theta \rangle$ & 
  $\langle \sigma_\theta\rangle$ & 
  $\sigma(Z_\theta)$ & 
  $\langle \Delta_\theta \rangle$ & 
  $\langle \sigma_\theta\rangle$ & 
  $\sigma(Z_\theta)$ & 
  $\langle \Delta_\theta \rangle$ & 
  $\langle \sigma_\theta\rangle$ & 
  $\sigma(Z_\theta)$ & 
  $\langle \Delta_\theta \rangle$ & 
  $\langle \sigma_\theta\rangle$ & 
  $\sigma(Z_\theta)$ \\

\hline
\hline
$f\sigma_8$ & 0.19 & 0.69 & 0.38 & 0.47 & 1.08 & 0.79 & 0.51 & 1.02 & 0.31 & 0.09 & 0.51 & 0.54 \\
\hline
\hline
$\omega_\textrm{m}$ & -0.02 & 0.95 & 0.27 & -0.12 & 0.90 & 0.41 & -0.20 & 0.93 & 0.31 & -0.07 & 0.93 & 0.43 \\ 
$\omega_\textrm{b}$ & -0.01 & 0.09 & 0.33 & -0.01 & 0.10 & 0.29 & -0.01 & 0.10 & 0.23 & -0.02 & 0.08 & 0.40 \\ 
$\sigma_8 $  & 0.56 & 2.06 & 0.35 & 1.44 & 2.16 & 0.54 & 0.66 & 2.98 & 0.40 & 1.05 & 1.60 & 0.47 \\ 
$w_0$  & -0.10 & 12.83 & 0.33 & 1.67 & 11.40 & 0.44 & -0.39 & 13.60 & 0.27 & 3.34 & 6.31 & 0.50 \\ 
$w_\textrm{a} $  & 0.40 & 32.77 & 0.30 & -1.38 & 32.96 & 0.44 & 0.52 & 32.57 & 0.29 & -2.52 & 28.36 & 0.32 \\ 
$h $ & 0.86 & 3.15 & 0.33 & 1.82 & 3.29 & 0.58 & 1.13 & 3.52 & 0.26 & 0.66 & 2.64 & 0.43 \\ 
$n_\textrm{s}$  & 0.87 & 2.87 & 0.41 & 0.80 & 2.86 & 0.44 & 0.42 & 3.08 & 0.23 & 0.68 & 2.19 & 0.51 \\ 
$N_\textrm{ur}$ & -0.43 & 52.12 & 0.21 & -3.26 & 57.55 & 0.22 & 3.44 & 56.77 & 0.22 & -4.68 & 50.45 & 0.28 \\ 
$\alpha_\textrm{s}$ & 0.31 & 1.70 & 0.28 & 0.52 & 1.85 & 0.52 & 0.16 & 2.12 & 0.22 & 0.11 & 1.43 & 0.56 \\ 
    
\hline
\hline 

$\alpha$ & 3.24 & 9.94 & 0.53 & 3.84 & 9.96 & 0.69 & 3.54 & 10.99 & 0.49 & 1.73 & 6.12 & 0.84 \\ 
$\kappa$ & -0.33 & 23.51 & 0.39 & -6.79 & 23.37 & 0.41 & -3.27 & 25.70 & 0.25 & -3.74 & 18.68 & 0.69 \\ 
$\log M_1$  & -2.31 & 6.98 & 0.55 & -1.63 & 8.38 & 0.57 & -2.31 & 7.97 & 0.62 & 0.43 & 3.94 & 0.61 \\ 
$\log M_\textrm{cut}$ & 0.35 & 6.24 & 0.40 & -0.35 & 8.77 & 0.58 & -0.18 & 8.05 & 0.30 & 1.13 & 3.99 & 0.62 \\ 
$\log \sigma$ & 0.79 & 6.80 & 0.43 & 1.42 & 9.51 & 0.49 & 1.94 & 10.24 & 0.35 & 1.24 & 5.00 & 0.60 \\ 

\hline
\hline
  
  \end{tabular}
  }
  \label{tab:merged_stats_ideal}
\end{table*}

We perform four different inferences on the 25 mocks of the recovery set:
galaxy auto-correlation only $\left[\xi_0^{gg},\xi_2^{gg} \right]$, velocity auto-correlation only $\left[\xi_0^{vv},\xi_2^{vv} \right]$, galaxy-velocity cross-correlations only $\left[\xi_1^{vg}\right]$ and all multipoles at once.

Table \ref{tab:merged_stats_ideal} presents the summary statistics computed from the 25 MCMC chains (the posterior distribution averaged over the different realisations is shown in Appendix \ref{annex:posterior}).
We find that averaging over the recovery set for every parameter $\theta$,
the difference between the expected and best fitted parameter value $\langle \Delta_\theta \rangle$
is well within the estimated one sigma $\langle \sigma_\theta\rangle$.
However, the standard deviation of the standard score $\sigma(Z_\theta)$ - with $Z_\theta \equiv \left(\theta - \langle \theta \rangle \right) / \sigma_\theta$ - 
is smaller than one for every parameter and every method.
While it is challenging to conclude from such a small sample,
this might indicate a systematic overestimation of the uncertainties, 
or a deviation from Gaussianity in the fitted parameter distributions. 
The reason could be a wrong estimation of the emulator error, as we have seen that the transformation Eq \ref{eq:slog} induces non-Gaussian predictions for the GP.
The mean reduced chi-squared are equal to 1.1 for the "gg" analysis, 
0.96 for "vv", 1.49 for "vg", and 0.85 for the complete joint analysis.
While those values deviates from one, the corresponding p-value $\left[\text{gg},\text{vv},\text{vg},\text{tot}\right]_\textrm{p-value} = \left[0.3,0.6,0.1,0.9\right]$
indicate reasonably good fits.

We find that the majority of the cosmological parameters are mainly constrained from the prior, as the estimated uncertainties are similar to the expected standard deviation of the uniform prior described in table \ref{tab:cosmo_hod_range}.
The parameters $\sigma_8$ and $w_0$ appear to be the most constrained by the measurements.
Compared to the inference using only the galaxy auto-correlation, 
using the velocity auto-correlation or the cross-correlation yield looser constraints on $\sigma_8$.
This is expected as the velocity perturbations are suppressed on smaller scales.
Nonetheless, velocity clustering on its own provides more information regarding $w_0$ than galaxy clustering alone.
Galaxy clustering is more constraining
for the HOD parameters, which only affect the velocity clustering on scales typically below 
$\sim 10 \hmpc$.
For almost every parameter, we obtain the tightest constraints when combining all clustering measurements.


\begin{figure}
  \centering

 \includegraphics[width=1.\columnwidth]{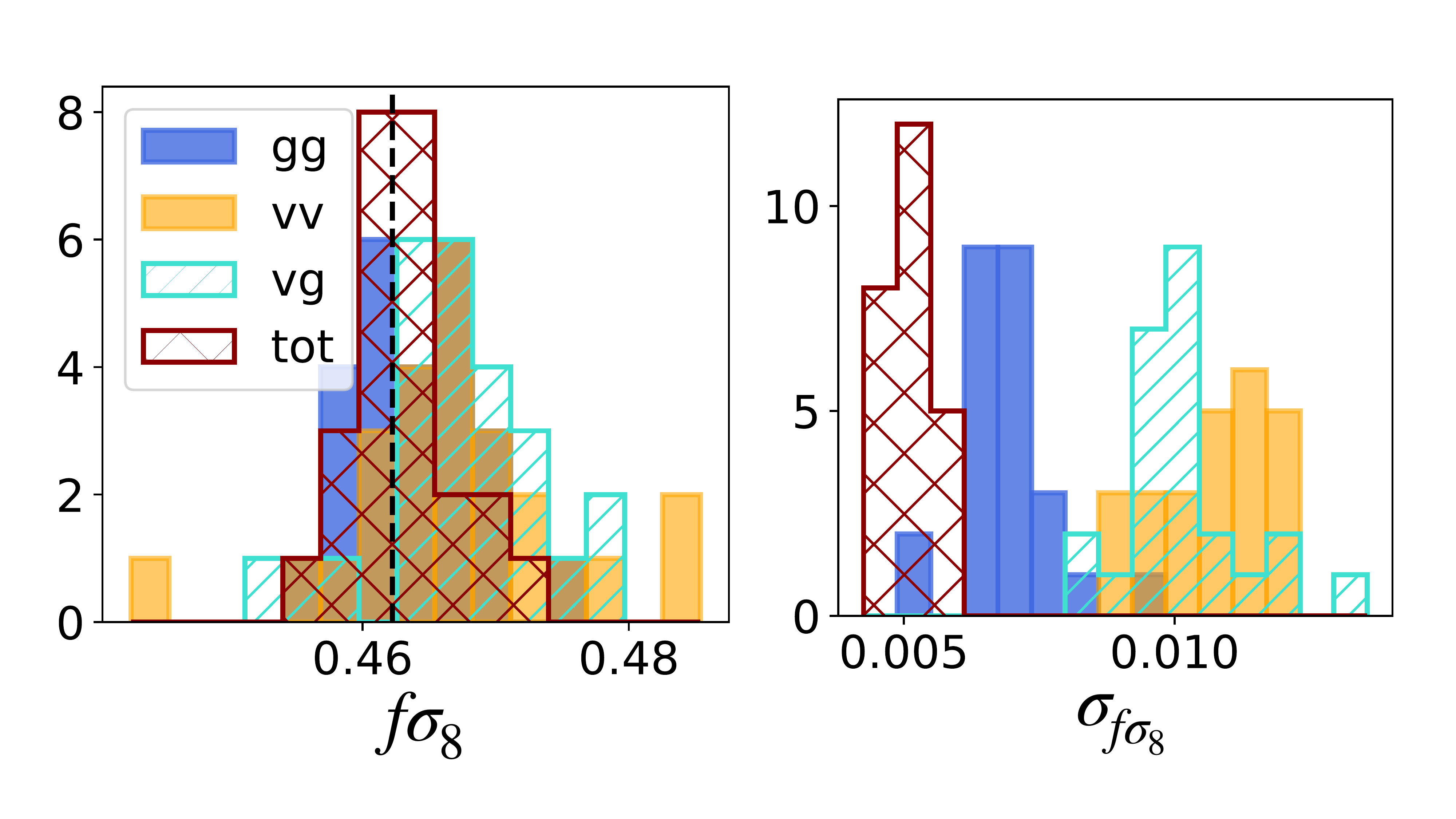} 

\caption{Growth rate best fit values along with the corresponding uncertainties derived from the cosmological parameters posteriors.
The results obtained using the galaxy, velocity, cross, and total data vectors are shown in blue, orange, cyan and red respectively. The expected growth rate value is pointed by a black dotted line. }
\label{fig:fs8}
\end{figure}

For every best fitted cosmological configuration, we derive the corresponding growth rate $f\sigma_8$ (at $z=0.2$). Its summary statistics is presented in table \ref{tab:merged_stats_ideal} and the corresponding distribution in figure \ref{fig:fs8}.
The results obtained using the galaxy, velocity, cross, and total data vectors are shown in blue, orange, cyan and red respectively.
We find that combining galaxy and peculiar velocity information improves the constraining power, yielding a $1.1\%$ precision measurement on $f\sigma_8$ compared to $1.5\%$ using galaxies alone.
While using the "gg", "vv", and "vg" methods
the estimated values are well within $1\sigma$, the systematic shifts are larger than the expected dispersion $\left ( \langle \sigma_\theta \rangle /\sqrt{25} \right )$.
Combining the different tracers reduces the average shift, making it consistent with the expected variation (although drawing conclusive results from 25 realisations is challenging).

We can evaluate the statistical significance 
on the total cosmological and HOD parameters biases,
accounting for the different correlations between parameters.
We compress the posterior information using the Figure of Merit (FoM)
and Figure of Bias (FoB) defined as 
\begin{equation}
\begin{aligned}
\mathrm{FoB}&=\sqrt{\left(\theta_\text{ best }-\theta_{\text {true }}\right)^{\top} C_{\theta}^{-1}\left(\theta_{\text {best }}-\theta_{\text {true }}\right)},
\\ \mathrm{FoM}&=\frac{1}{\sqrt{ |C_{\theta}|}},
\end{aligned}
\label{eq:fom_fob}
\end{equation}
where $C_{\theta}$ is the parameter covariance matrix estimated from the posterior distribution.
Assuming that a given set of parameters $\theta$ are Gaussianly distributed (following the posterior),
the FoB is the square root of the corresponding chi-squared evaluated at the parameters best-fit values.
This metric indicates how significant the biases on the parameters are (assuming a Gaussian posterior).
The FoM is useful to estimate the overall constraints on a set of parameters, accounting for the correlations.
For every realisation of the recovery set, we compute the FoM and FoB for the 9 cosmological parameters' subspace $\theta_\textrm{cosmo}$, 
and the 5 HOD parameters sub-spaces $\theta_\textrm{HOD}$.

\begin{figure}
  \centering

 \includegraphics[width=0.9\columnwidth]{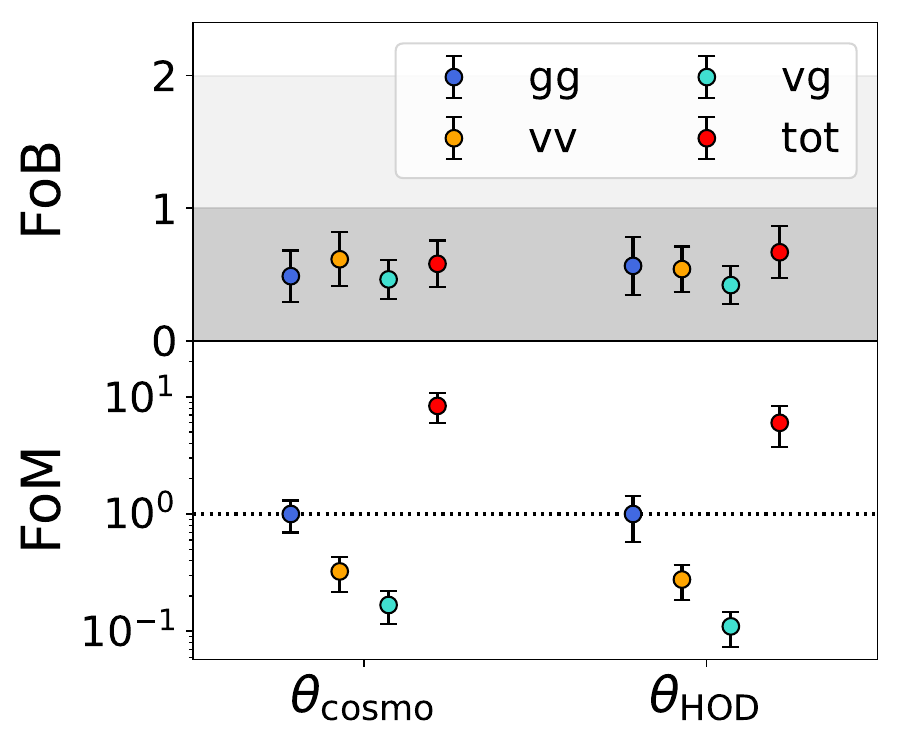} 

\caption{Cosmological and HOD parameters recovery for the four inference strategies run on the recovery set.
Top panel : FoB normalised by the $\sqrt{\chi^2}$ associated with a p-value of $0.68\%$, that is 10.4 for $\theta_\textrm{cosmo}$
and 5.9 for $\theta_\textrm{HOD}$. The shaded area correspond to 1 and 2 $\sigma$ CL.
Bottom panel : FoM normalised by the "gg" values : $4.97 \times 10^{14}$ for $\theta_\textrm{cosmo}$
and $2.20 \times 10^{6}$ for $\theta_\textrm{HOD}$.}
\label{fig:fom_ideal}
\end{figure}

Figure \ref{fig:fom_ideal} gives the average results for the corresponding "gg", "vv", "vg" and "tot" analyses, designed by the same colors as previously.
The errorbars correspond to the standard deviations of the FoM and FoB across the 25 realisations.
The bottom panel shows the cosmological and HOD mean FoMs, normalised by the value obtained for the "gg" analysis : $4.97 \times 10^{14}$ for $\theta_\textrm{cosmo}$
and $2.20 \times 10^{6}$ for $\theta_\textrm{HOD}$.
The top panel of figure \ref{fig:fom_ideal} shows the FoB normalised by the $\sqrt{\chi^2}$ values associated with a p-value of $0.68\%$,
corresponding to $\sqrt{\chi^2} = 10.4$ for the 9 dof cosmological space, and $\sqrt{\chi^2} = 5.9$ for the 5 dof HOD space.
Hence, the grey shaded area correspond to 1 and 2 $\sigma$ CL.
We find unbiased $\theta_\textrm{cosmo}$ and $\theta_\textrm{HOD}$ for every inference strategy, 
and retrieve the same ordering of the constraining power $\text{FoM}_\textrm{vg}<\text{FoM}_\textrm{vv}<\text{FoM}_\textrm{gg}<\text{FoM}_\textrm{tot}$.
On mildly to non-linear scales, peculiar velocity clustering holds additional information not contained in the galaxy clustering alone,
both for the cosmological and galaxy-halo bias parameters.

\section{Inference on noisy measurements}
\label{sec:cosmology_constraints} 

So far we have assumed that every galaxy had its peculiar velocity perfectly measured without any noise.
In this section we study how constraints are impacted when using sparse noisy measurements of peculiar velocities.

\subsection{ Constraining power}

Following the same inference strategy (with the corresponding data covariance matrix), 
we fit the clustering of the 25 realistic mocks described in section \ref{sec:data:realistic}, 
reproducing the expected density of tracer, uncertainties on the velocity measurements, and associated weights (Eq \ref{eq:weights_fkp}).
Additionally, we include AP distortions in the clustering, using the \textsc{c003} fiducial cosmology in \textsc{AbacusSummit}
to convert redshifts into comoving distances.
Given that we do not have a theoretical prediction for the 2D correlation functions $\xi^{ab}(s,\mu)$, we model the AP distortions
 - using the isotropic and anisotropic dilation's parameters $(\alpha_V,\epsilon)$ -
directly on the multipoles, following \cite{xu_measuring_2013} at first order (neglecting the derivative terms).
The parameters are updated for each likelihood evaluation.
Note that an additional dependence on the fiducial cosmology is also introduced in the velocity measurement through Eq \ref{eq:eta_to_v}.
We leave the detailed study of this effect to future work and use the true underlying cosmology (Planck18) to compute 
velocities from log-distance ratios.

\begin{table*}
  \centering
  \caption{Summary statistics for the fits of the 25 realistic mocks with Planck2018 cosmology 
  and the median HOD. $\langle \Delta_\theta \rangle$ is the bias, the difference between the true and best fitted value for any parameter $\theta$ scaled by $10^2$.
   $\langle \sigma_\theta\rangle$ is the average over the 25 realisations of the (symmetrized) one sigma confidence level scaled by $10^2$. 
   $\sigma(Z_\theta)$ is the standard deviation of the standard score for any parameter $\theta$. 
   Note that $f\sigma_8$ is a derived parameter from the other cosmological ones, and corresponds to redshift $z=0.2$.}
   \small
   \addtolength{\tabcolsep}{-2pt}
  {
  \begin{tabular}{l|ccc|ccc|ccc|ccc}
  \hline
  \hline
     & \multicolumn{3}{c|}{gg }  & \multicolumn{3}{c|}{vv } & \multicolumn{3}{c|}{vg } & \multicolumn{3}{c}{tot }  \\
   $p$ &
  $\langle \Delta_\theta \rangle$ & 
  $\langle \sigma_\theta\rangle$ & 
  $\sigma(Z_\theta)$ &
  $\langle \Delta_\theta \rangle$ & 
  $\langle \sigma_\theta\rangle$ & 
  $\sigma(Z_\theta)$ &
  $\langle \Delta_\theta \rangle$ & 
  $\langle \sigma_\theta\rangle$ & 
  $\sigma(Z_\theta)$ &
  $\langle \Delta_\theta \rangle$ & 
  $\langle \sigma_\theta\rangle$ & 
  $\sigma(Z_\theta)$ \\

  \hline
  \hline
  $f\sigma_8$ &  0.89 & 2.16  & 0.55 & 1.90 & 4.67  & 0.18 & 5.22 & 4.16  & 0.38 & 1.32 & 1.78 & 0.61 \\ 
  \hline
  \hline
  $\omega_\textrm{m}$ & -0.09 & 0.89 & 0.60 & -0.44 & 1.12 & 0.21 & -0.47 & 0.97 & 0.40 & -0.14 & 0.87 & 0.49 \\ 
  $\omega_\textrm{b}$ & -0.00 & 0.09 & 0.54 & -0.02 & 0.11 & 0.29 & -0.02 & 0.10 & 0.34 & -0.01 & 0.09 & 0.54 \\ 
  $\sigma_8 $  & -1.64 & 4.16 & 0.44 & -0.43 & 5.97 & 0.53 & -5.20 & 5.46 & 0.37 & -1.93 & 3.14 & 0.64 \\ 
  $w_0$  & -2.62 & 14.12 & 0.43 & -0.19 & 16.83 & 0.17 & 1.17 & 15.45 & 0.46 & -0.26 & 12.48 & 0.52 \\ 
  $w_\textrm{a} $  & -4.03 & 31.38 & 0.60 & 2.13 & 37.36 & 0.30 & 2.01 & 36.76 & 0.31 & -3.68 & 30.80 & 0.53 \\ 
  $h $ & 0.26 & 3.43 & 0.51 & 2.48 & 4.78 & 0.22 & 3.43 & 4.60 & 0.39 & 0.20 & 2.89 & 0.53 \\ 
  $n_\textrm{s}$  & 0.66 & 3.07 & 0.47 & -0.64 & 3.55 & 0.18 & -0.15 & 3.40 & 0.38 & 0.90 & 2.83 & 0.50 \\ 
  $N_\textrm{ur}$ & 5.43 & 51.68 & 0.45 & -4.64 & 61.66 & 0.26 & -1.32 & 58.10 & 0.34 & 13.51 & 47.56 & 0.63 \\ 
  $\alpha_\textrm{s}$ & 0.44 & 1.86 & 0.50 & -0.04 & 2.31 & 0.24 & -0.30 & 2.10 & 0.37 & 0.10 & 1.83 & 0.57 \\ 

  \hline
  \hline 

  $\alpha$ & -31.07 & 13.82 & 0.46 & -8.78 & 37.02 & 0.34 & -16.31 & 34.77 & 0.30 & -33.60 & 13.85 & 0.64 \\ 
  $\kappa$ & -2.76 & 23.17 & 0.48 & 1.54 & 29.76 & 0.29 & 3.96 & 29.66 & 0.27 & -3.29 & 22.67 & 0.58 \\ 
  $\log M_1$  & 51.93 & 7.58 & 0.38 & 32.74 & 29.90 & 0.27 & 41.58 & 22.82 & 0.30 & 53.39 & 7.18 & 0.46 \\ 
  $\log M_\textrm{cut}$ & 22.89 & 20.96 & 0.47 & 15.25 & 32.93 & 0.22 & 19.35 & 33.39 & 0.28 & 27.03 & 21.35 & 0.57 \\ 
  $\log \sigma$ & -25.73 & 8.32 & 0.56 & -14.40 & 25.78 & 0.31 & -20.74 & 23.13 & 0.31 & -24.21 & 9.50 & 0.60 \\ 
 
  \hline
  \hline
  \end{tabular}
  }
  \label{tab:stats_pvmocks}
\end{table*}

Table \ref{tab:stats_pvmocks} presents the summary statistics computed from the fits.
We find similar results to the previous section.
While the standard deviation of the standard score $\sigma(Z_\theta)$ is small compared to one, each of the cosmological parameters are recovered within 1$\sigma$.
The "vv" and "vg" analyses yield larger systematic shift and errorbars compared to the "gg" results,
and the combined probes give the tightest (unbiased) cosmological constraints.
We get the following mean reduced chi-squared $\left[\text{gg},\text{vv},\text{vg},\text{tot}\right]_{_r\chi^2} = \left[1.1,4.3,3.0,0.4\right]$,
with the corresponding p-values $\left[\text{gg},\text{vv},\text{vg},\text{tot}\right]_\textrm{p-value} = \left[0.3,2\times 10^{-20},6 \times 10^{-5}, 1 - 1 \times 10^{-10}\right]$.
While the AP distortion modelling is only approximated, 
it remains a weak effect and is expected to cause the worse quality fits for the "vv" and "vg" inferences, 
most likely due to the reduced density and the random scatter in the velocities.
As seen in figure \ref{fig:pv_noise_estimator}, while the random noise in the velocity measurement is almost averaged out for the largest scales,
on the smallest scales it reduces the correlations between separation bins and significantly increases the variance of the clustering.
For the HOD parameters which strongly affect those scales, we find systematic biases comparable or larger than the estimated uncertainties.

Once again, for every cosmological parameter configurations, we derive the corresponding growth rate distribution and give its statistics in table \ref{tab:stats_pvmocks}. 
While the statistical gain is less significant than using the previous ideal measurements, we still find an improvement when combining galaxy with peculiar velocity clustering, 
reaching $3.8\%$ precision measurement on $f\sigma_8$ compared to $4.7\%$ for galaxy clustering alone.
However, measurements using the galaxy-velocity cross correlation yields $f\sigma_8$ estimates biased by more than $1\sigma$.
While the joint analysis of galaxy and velocity clustering yields unbiased constraints, 
the resulting systematic shift is larger than the expected dispersion of the average measurement.
Although the amount of available realisations is too small to draw significative conclusions,
those results point toward a potential systematic bias that could cancel out the statistical gain of including velocity clustering on the non-linear scales.

\begin{figure}
  \centering

 \includegraphics[width=0.9\columnwidth]{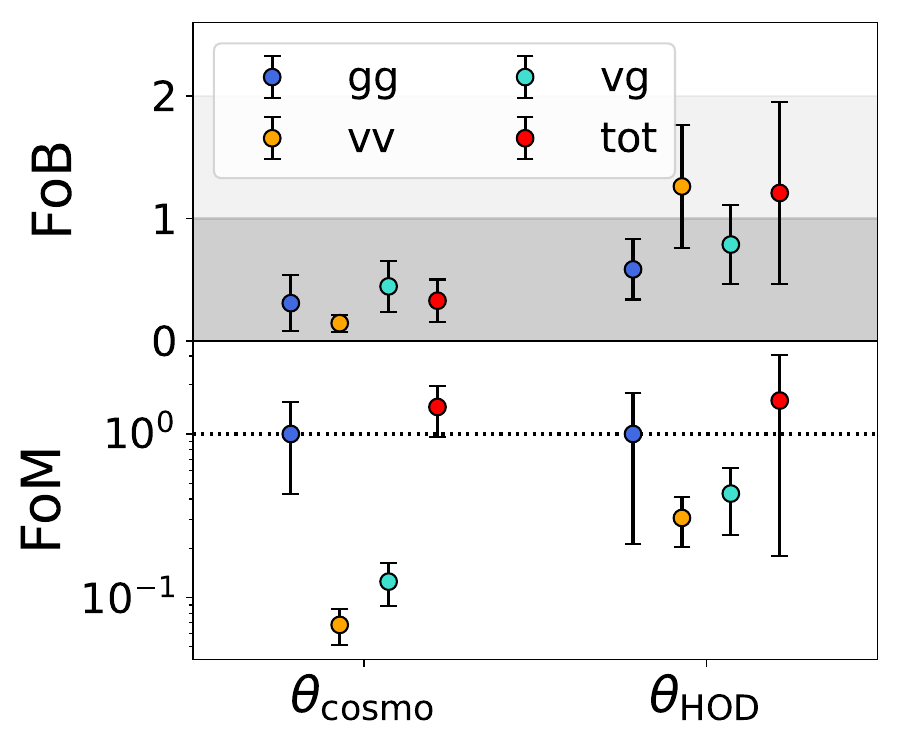} 

\caption{Cosmological and HOD parameters recovery for the four inference strategies run on the realistic mocks.
Top panel : FoB normalised by the $\sqrt{\chi^2}$ associated with a p-value of $0.68\%$, that is 10.4 for $\theta_\textrm{cosmo}$
and 5.9 for $\theta_\textrm{HOD}$. The shaded area correspond to 1 and 2 $\sigma$ CL.
Bottom panel : FoM normalised by the "gg" values, $4.85 \times 10^{13}$ for $\theta_\textrm{cosmo}$
and $5.59 \times 10^{4}$ for $\theta_\textrm{HOD}$.}
\label{fig:fom_pvmocks}
\end{figure}

As done previously, in order to assess the overall systematic bias and constraining power,
 we compute for each analysis the mean and standard deviation of the FoM and FoB in the parameter subspaces $\theta_\textrm{cosmo}$ 
and $\theta_\textrm{HOD}$.
Figure \ref{fig:fom_pvmocks} presents the results obtained for the four methods, preserving the same colour scheme.
First, we notice that fitting only the velocity-autocorrelation or the total joint clustering yields a $1\sigma$ level bias on HOD parameters.
The FoB of the cross-correlation analysis is increased as well, but remains on average bellow one.
Furthermore, we notice that compared to the noise-free high-density constraints in figure \ref{fig:fom_ideal},
the FoM for the gg analysis - used as a normalisation - dropped by a factor 10 for $\theta_\textrm{cosmo}$, and almost 40 for $\theta_\textrm{HOD}$,
indicating a loss of constraining power.
Yet, for every method we obtain unbiased cosmological parameters.
Despite the reduced number of tracers and the noisy peculiar velocity measurements, 
we find that jointly fitting the galaxy and velocity clustering improves the cosmological constraints, 
even though the improvement is less significant than when using the full velocity sample.

%

\subsection{Varying the minimum fitting scale}

\begin{figure}
  \centering

 \includegraphics[width=0.9\columnwidth]{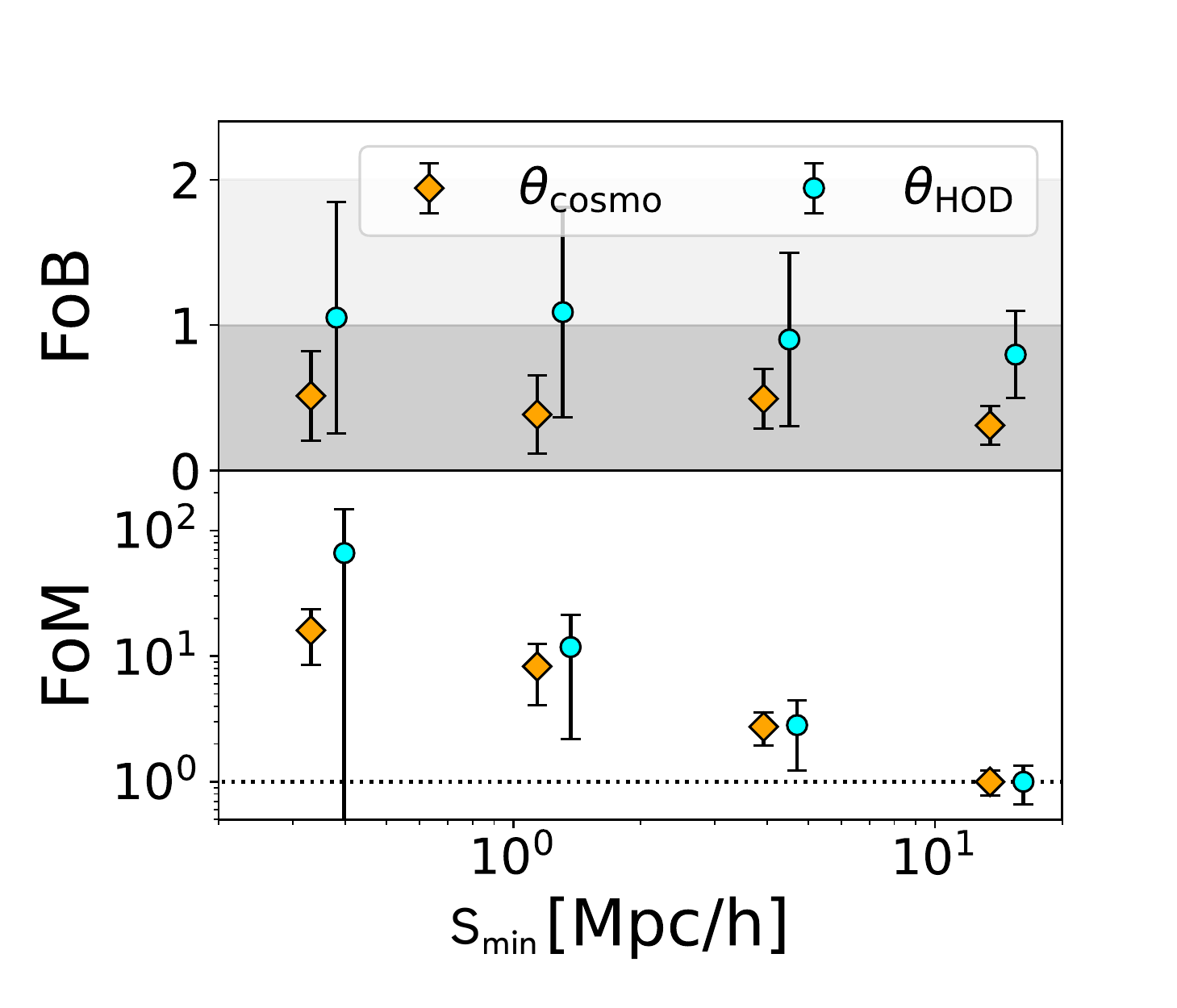} 

\caption{Cosmological and HOD parameters recovery using different minimum scales of the realistic mocks to run inference with the total clustering information.
Top panel : FoB normalised by the $\sqrt{\chi^2}$ associated with a p-value of $0.68\%$, that is 10.4 for $\theta_\textrm{cosmo}$
and 5.9 for $\theta_\textrm{HOD}$. The shaded area correspond to 1 and 2 $\sigma$ CL.
Bottom panel : FoM normalised by the value obtained for the largest $s_\textrm{min}$, $5.96 \times 10^{12}$ for $\theta_\textrm{cosmo}$
and $1.32 \times 10^{3}$ for $\theta_\textrm{HOD}$.}

\label{fig:fom_scales}
\end{figure}

Since the noise in the peculiar velocity measurements mostly affects the smallest scales, 
the inference might be optimised by removing some separation bins, thereby further compressing the data vector.
We investigate the dependence of the constraining power and systematic bias on the minimum fitted scale.
We fit the total clustering information $\left[\xi_0^{gg},\xi_2^{gg},\xi_0^{vv},\xi_2^{vv},\xi_1^{vg} \right]$
for each individual realisations of the the realistic mocks, using different minimum scales 
$s_\textrm{min} = \left[0.33,1.13,3.92,13.4\right]$ in $\hmpc$, corresponding to $\left[30,21,14,7\right]$ data-points for each multipole.
We further compute for each $s_\textrm{min}$ the mean and standard deviation of the FoM and FoB for the cosmological and HOD parameters,
and show the results in figure \ref{fig:fom_scales}.

For the HOD parameters, we find that the mean FoM increases as we consider smaller scales, but its variance also rises. 
Conversely, while decreasing $s_\textrm{min}$ reduces uncertainties in the HOD parameters, it does not sufficiently mitigate the systematic shift, 
 leading to biased constraints with the mean FoB exceeding 1 for $s_\textrm{min} \gtrsim 4\hmpc$. 
However, the flexibility of the HOD formalism allows to marginalise over unmodelled noise, preventing bias in the cosmological parameters. 
For any minimum scale, the cosmological parameters FoB remains below 1.
Additionally, as the minimum fitted scale decreases, the cosmological parameters' FoM increases, tightening the constraints. 
We measure poor p-values, larger than $1-1 \times 10^{-3}$, except for the largest minimum scale $s_\textrm{min} = 13.4\hmpc$, where we find a p-value of 0.1.
This is in line with the fact that despite the enlarged variance on small scales, the HOD parameters are effectively overfitting the signal.

It is worth emphasising that for a real survey, fibre collision systematics could influence this result.
Indeed, if not properly corrected, fibre collisions mainly bias the clustering measurement for the smallest separation bins.
Additionally, while we found that the HOD parameters were able to absorb additional sources of noise, effectively unmodelled physics, 
our noisy sample is built from the same underlying formalism.
We acknowledge that this empirical relationship does not fully capture the complexity of galaxy formation physics. 
Therefore, before applying such emulator model to real data, it is crucial to assess its robustness against more
physically motivated galaxy-halo relation.
such as abundance matching technique (\cite{Kravtsov_AM_2004}), semi-analytical models of galaxy formation (\cite{White_SAM_1991}) 
hybrid bias expansion(\cite{Pellejero_hybridBias_2022}), or
hydrodynamical simulations (\cite{Vogelsberger_cosmosim_2020} for an overview).

\section{Conclusions}

In this work, we applied our estimator for the velocity two-point statistics within a simulation-based emulation framework.
 We trained emulator models for the redshift space galaxy and velocity 
clustering using the \textsc{AbacusSummit} simulations and HOD formalism. We demonstrated that using the galaxy-galaxy, 
velocity-velocity, or velocity-galaxy correlation functions, we were able to recover unbiased constraints for the cosmological and HOD parameters.
 Moreover, combining the velocity and galaxy clustering on mildly to non-linear scales yielded tighter constraints.

We built a set of realistic mocks, describing the typical density and error in the peculiar velocity measurements of 
type-Ia supernovae and the TF/FP relations corresponding to the ZTF and DESI surveys. 
While fitting those mocks results in systematic errors in HOD parameters, 
the cosmological parameters remain unbiased. 
The HOD model is flexible enough to marginalise over the noisy 
clustering without biasing the cosmological parameters.
Hence, the physical interpretation of the HOD parameters must be treated with caution.
Even with realistic measurements, we found that the joint analysis of galaxy and peculiar velocity clustering leads to tighter constraints.
Although the gain is reduced compared to the case where we have a velocity measurement for every galaxy, this motivates the measurement of 
larger velocity samples and reduction of uncertainties arising from distance indicator intrinsic scatter. Additionally, we found that even the smallest scales, 
going down to 0.33 $\hmpc$, improve the cosmological constraints.
Thus, it would be worthwhile to push the effort to correct for fibre collision systematics.
Using the full available scales, we found that combining galaxy and velocity tracers yields $3.8\%$ precision measurement on $f\sigma_8$, compared to $4.7\%$ using only the galaxy information.
However, including the velocity information tend to increase the systematic shift on $f\sigma_8$ which might reduce the importance of this statistical gain.

It is essential to ensure that the HOD parameters are properly sampled to describe the clustering of a given tracer. 
Making use of the MKGP formalism, the emulator parameter space could be extended to more elaborate HOD models, 
including assembly bias (\cite{Xu_AssBias_2021}) or velocity bias extension (\cite{Guo_VelBias_2015}) to describe any additional scatter. 
The emulator could also be extended to different redshift bins, additional multipoles, larger scales, or forward-modelled observational effects. 
To properly account for wide-angle effects for low redshift surveys, the emulated $\xi(s,\mu)$ (or multipoles) should either be trained on a given footprint 
or binned along an additional dimension to fully characterise triangles formed by each pair of galaxies and the observer. 
This last option would require a detailed study of the compression/information-loss balance.

It is worth mentioning that any emulator model highly depends on the simulation suite's numerical specificity and the galaxy model employed. 
In particular, the HOD is an empirical formalism and does not capture the true complexity of galaxy formation. 
Therefore, any emulator should be extensively tested against different simulations with various galaxy models before being applied to real data.

\subsection*{Acknowledgements}
The project leading to this publication has received funding from Excellence Initiative of Aix-Marseille University - A*MIDEX, a French ``Investissements d'Avenir'' program (AMX-20-CE-02 - DARKUNI). 
This work is supported by the U.S. Department of Energy, Office of Science, Office of High-Energy Physics, under Contract No. DE–AC02–05CH11231, and by the National Energy Research Scientific Computing Center. Additional support for DESI was provided by the U.S. National Science Foundation, Division of Astronomical Sciences under Contract No. AST-0950945; the Science and Technology Facilities Council of the United Kingdom; the Gordon and Betty Moore Foundation; the Heising-Simons Foundation; the French Alternative Energies and Atomic Energy Commission; the National Council of Humanities, Science and Technology of Mexico; the Ministry of Science, Innovation and Universities of Spain, and by the DESI Member Institutions. 
The authors are honored to conduct research on I'oligam Du'ag, a mountain particularly significant to the Tohono O’odham Nation.
For more information, visit desi.lbl.gov.


\subsection*{Data availability}
All data necessary to reproduce the analyses presented in
this paper are publicly available in a machine-readable form on \href{https://zenodo.org/records/18455435?token=eyJhbGciOiJIUzUxMiJ9.eyJpZCI6IjVlZDcwZGRkLTEyOTktNDQyMi04NDgzLWU4Y2FkYjU2MzQ1ZSIsImRhdGEiOnt9LCJyYW5kb20iOiJlMDZkM2VjYzllNWUyZDdlYmU3OTJmYzg4ZTNhNWRhNyJ9.j6uWhjnPB9JNX56lbAQRGK0DlHBb7y_FdNuY2mfoTTNO5fgxM8WR98JCH3Sil2oZHu7xjWZzvOCXsEreBeh_4Q}{Zenodo}
%

\bibliographystyle{aa}
\bibliography{Biblio,DESI_PV_DR1}

\newpage
\appendix

\section{Correlation function estimator}
\label{sec:annex:estimator}
We give a complete derivation for the state of the art Landy-Szalay estimator \citep{landy-szalay} for the galaxy correlation function, and use the same methodology to derive estimators for the velocity auto and cross correlation functions. \\
The correlation function of two random fields $\delta_{\rm a}$, $\delta_{\rm b}$ is defined as
\begin{equation}
\xi_{\rm ab}(\textbf{x}_1-\textbf{x}_2) \equiv \langle \delta_{\rm a}(\textbf{x}_1)\delta_{\rm b}(\textbf{x}_2) \rangle,
\label{eq:TPCF}
\end{equation}
For the clustering of galaxy, we want to estimate the over density of galaxy $\delta_g(\textbf{x})$. 
The galaxy spatial distribution is obtained by running a discrete sum $n_g(\mathbf{x}) = \sum_i^{\rm N_g} \delta^{\rm D}(\mathbf{x} - \mathbf{x_i}) \ w_{g,i} $, with $\delta^{\rm D}$ the Dirac delta function $N_g$ the number of observed galaxies, and $w_{g,i}$ the weight associated with a given galaxy $i$ to correct for systematic effect. For a complete sample without any observational effect the weights are set to $1$.
In order to derive and estimator for the galaxy correlation function, we can also write the observed galaxy density as  
\begin{equation}
n_g(\mathbf{x}) = \Bar{n}_g \mathcal{W}_g (\mathbf{x}) \left( 1 + \delta_{\rm g}(\mathbf{x}) \right) \frac{1}{w_g(\mathbf{x})},
\end{equation}
with $\Bar{n}_g$ the mean density of galaxies and $\mathcal{W}_g$ the window function describing the survey footprint.
A random catalogue $n_s(\mathbf{x})$ following the spatial distribution of galaxies is commonly used to describe the product $\Bar{n}_g\mathcal{W}_g(\mathbf{x})$. A number of random sample $N_s$ a few ten times larger than $N_g$ is usually enough to properly sample the footprint.
The random catalogue can also have associated weights $w_s(\mathbf{x})$ to reproduce observational systematics.
Rewriting the previous relation we define the galaxy overdensity field as
\begin{equation}
\delta_{\rm g}(\mathbf{x}) = \frac{w_g(\mathbf{x}) n_g(\mathbf{x}) - w_s(\mathbf{x}) n_s(\mathbf{x})}{w_s(\mathbf{x}) n_s(\mathbf{x})}.
\label{eq:delta_g}
\end{equation}
Following Eq \ref{eq:TPCF}, we square the field and assume ergodicity so the ensemble averaging $\langle . \rangle $ is equivalent to a spatial averaging.
Doing so we can derive the well known Landy-Szalay estimator \citep{landy-szalay} for the galaxy TPCF 

\begin{equation}
\xi_{gg}(\textbf{r}) = \frac{DD(\textbf{r}) -DS(\textbf{r}) - SD(\textbf{r}) }{SS(\textbf{r})} + 1,
\label{eq:LS}
\end{equation}
where $\textbf{r} = (\textbf{x}_1 - \textbf{x}_2)$ and $DD$, $RR$, $DR$ and $RD$ are the normalised auto and cross pair counts of the data and random catalogues given by
\begin{equation}
\begin{aligned}
&DD(\textbf{r}) = \langle n_g(\textbf{x}_1)n_g(\textbf{x}_2) \rangle =  \frac{1}{N_{\rm  gg}} \sum_i^{\rm N_g} \sum_{j \neq i}^{\rm N_g} \delta^{\rm D}(\mathbf{r} - |\mathbf{x}_{g,i} - \mathbf{x}_{g,j}|) \ w_{g,i} w_{g,j},\\
&SS(\textbf{r}) =  \langle n_s(\textbf{x}_1)n_s(\textbf{x}_2) \rangle =  \frac{1}{N_{\rm  ss}} \sum_i^{\rm N_s} \sum_{j \neq i}^{\rm N_s} \delta^{\rm D}(\mathbf{r} - |\mathbf{x}_{s,i} - \mathbf{x}_{s,j}|) \ w_{s,i} w_{s,j},\\
&DS(\textbf{r}) = \langle n_g(\textbf{x}_1)n_s(\textbf{x}_2) \rangle =  \frac{1}{N_{\rm  gs}} \sum_i^{\rm N_g} \sum_{j }^{\rm N_s} \delta^{\rm D}(\mathbf{r} - |\mathbf{x}_{g,i} - \mathbf{x}_{s,j}|) \ w_{g,i} w_{s,j},\\
&SD(\textbf{r}) = \langle n_s(\textbf{x}_1)n_g(\textbf{x}_2) \rangle =  \frac{1}{N_{\rm  sg}} \sum_i^{\rm N_s} \sum_{j }^{\rm N_g} \delta^{\rm D}(\mathbf{r} - |\mathbf{x}_{s,i} - \mathbf{x}_{g,j}|) \ w_{s,i} w_{g,j}.
\end{aligned}
\label{eq:pair_counts}
\end{equation}
The auto and cross normalisation factors for two catalogues $a$ and $b$ are defined as
\begin{equation}
\begin{aligned}
&N_{\rm aa} =  \left (\sum_i^{\rm N_a} w_{a,i}\right )^2 - \sum_i^{\rm N_a} w_{a,i}^2,\\
&N_{\rm ab} =  \left (\sum_i^{\rm N_a} w_{a,i}\right ) \left (\sum_j^{\rm N_b} w_{b,j}\right ).
\end{aligned}
\label{eq:pair_counts_norm}
\end{equation}

Following the same methodology we can derive Landy-Szalay-like estimators for the radial velocity auto-correlation and the galaxy-velocity cross correlation. 
As in large-scales surveys only the radial component of the galaxy velocity field can be measured, in the following any use of "velocity" implies "radial velocity".

As the peculiar velocities are measured only where galaxies are observed, the field sampled is the galaxy momentum field $p_g(\textbf{x})$. Its spatial distribution can be written as a discrete sum $p_g(\mathbf{x}) = \sum_i^{\rm N_v} \delta^{\rm D}(\mathbf{x} - \mathbf{x}_i) \ u_{g,i} \ w_{v,i} $, with $ N_v$ the number of peculiar velocity measurements, $u_{g,i}$ and $w_{v,i}$ the velocity and weight respectively evaluated at position $\textbf{x}_i$.
The galaxy momentum can also be expressed as
\begin{equation}
p_g(\mathbf{x}) = \Bar{n}_v \mathcal{W}_v (\mathbf{x})  \ u_g(\mathbf{x}) \ \left( 1 + \delta_{\rm g}(\mathbf{x}) \right) \  \frac{1}{w_v(\mathbf{x})},
\end{equation}
with $\Bar{n}_v$ the mean density of velocity measurements and $\mathcal{W}_v$ the window function describing the peculiar velocity survey footprint. Again, we can use a catalogue $n_r(\textbf{x})$ of $N_r$ random sample to describe $\Bar{n}_v \mathcal{W}_v(\textbf{x})$. 
We call the weights associated with the random catalogue $w_r(\textbf{x})$.
We rewrite the previous relation to \textbf{isolate the fields of interest from the selection function} and introduce the galaxy weighted velocity field as
\begin{equation}
v(\textbf{x}) = u_g(\mathbf{x}) \ \left( 1 + \delta_{\rm g}(\mathbf{x}) \right) = \frac{w_v(\mathbf{x}) p_g(\mathbf{x})}{w_r(\mathbf{x}) n_r(\mathbf{x})}.
\label{eq:v_g}
\end{equation}
Once again, following Eq \ref{eq:TPCF} we can define the auto and cross correlation function by respectively squaring Eq \ref{eq:v_g} and multiplying it by Eq \ref{eq:delta_g} :
\begin{equation}
\xi_{vv}(\textbf{r}) = \frac{VV(\textbf{r}) }{RR(\textbf{r})},
\label{eq:LS_vv}
\end{equation}
\begin{equation}
\xi_{vg}(\textbf{r}) = \frac{VD(\textbf{r}) - VS(\textbf{r}) }{RS(\textbf{r})},
\label{eq:LS_vg}
\end{equation}
Following the same convention for the pair-counts and the normalisation. Note that $\xi_{gv}$ encapsulates as much information as $\xi_{vg}$ and therefore is not used here.
In the following we refer to the "galaxy weighted velocity field"  as the "velocity field" since there are equivalent in linear theory. However the reader should keep in mind that according to our definition, the velocity correlation function written as 

$$\xi_{vv}(\textbf{r}) = \langle u_g(\mathbf{x}_1) \ \left( 1 + \delta_{\rm g}(\mathbf{x}_1) \right) u_g(\mathbf{x}_2) \ \left( 1 + \delta_{\rm g}(\mathbf{x}_2) \right) \rangle$$
should encapsulate on non-linear scales the information contained in $\xi_{gg}$, $\xi_{uu}$,$\xi_{ug}$ and higher order statistics.

\section{Peculiar velocity from distance indicator}
\label{annex:vel}
Once the log-distance ratio is known, deriving the peculiar velocity is straightforward.
From the Doppler shift in the spectroscopic redshift
\begin{equation}
    1 + z_\textrm{obs} = (1 + z_\textrm{cos})(1 + v/c),
\label{eq:rsd}
\end{equation}
the comoving radial distance can be expanded at linear order in $v/c$ 
\begin{equation}
    D(z_\textrm{ cos}) = D(z_\textrm{ obs}) + \frac{c}{H(z_\textrm{ obs})} \left( z_\textrm{ cos} - z_\textrm{ obs}\right),
\end{equation}
leading to the distance ratio
\begin{equation}
    \frac{D(z_\textrm{ cos})}{D(z_\textrm{ obs})} = 1 - \frac{(1+ z_\textrm{ obs})}{D(z_\textrm{ obs})H(z_\textrm{ obs})} v.
\end{equation}
Finally, we can write the comoving radial distance ratio as
\begin{equation}
    \begin{aligned}
        \eta &\equiv - \log \frac{D(z_\textrm{ cos})}{D(z_\textrm{ obs})}\\
        &= - \log\left[1 - \frac{(1+ z_\textrm{ obs})}{D(z_\textrm{ obs})H(z_\textrm{ obs})} v \right]\\
        & = \alpha(z_\textrm{ obs}) v,
    \end{aligned}
\label{eq:eta_to_v}
\end{equation}
where the last equality is only valid at low redshift. 
The proportionality factor is defined as 
\begin{equation}
    \alpha(z_\textrm{ obs}) \equiv \frac{1}{\ln 10}  \frac{(1+ z_\textrm{ obs})}{D(z_\textrm{ obs})H(z_\textrm{ obs})}.
\label{eq:alpha_pv_factor}
\end{equation}
Note that this factor can slightly vary if measuring the luminosity or angular distance.

\section{Recovery set posterior distribution}
\label{annex:posterior}

\begin{figure*}
	\centering
  \includegraphics[width=1.\textwidth]{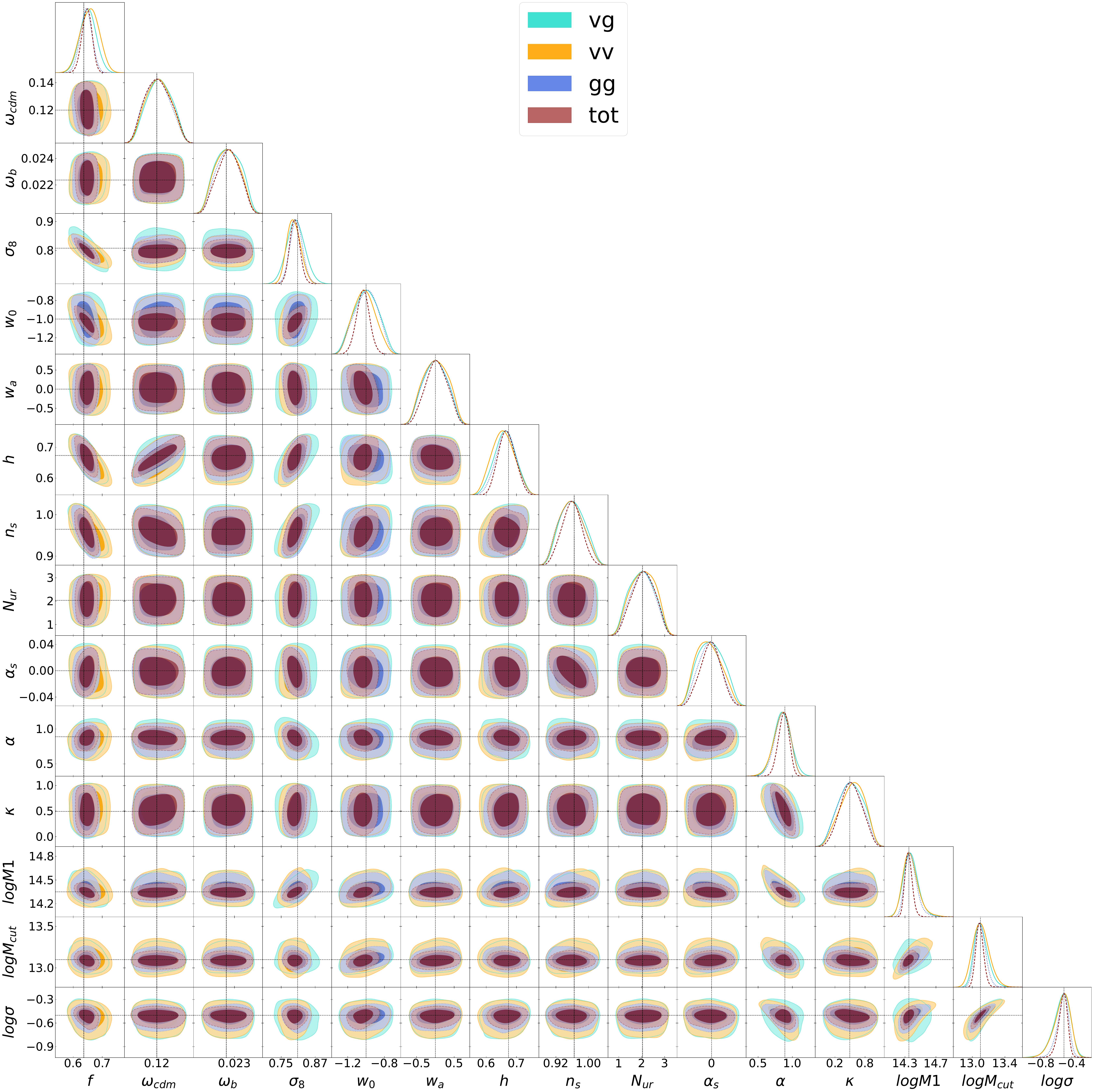}%
	\caption{Average posterior distribution of the inferred  parameters, from the fits of the recovery set.
  The contours correspond to 1 and 2$\sigma$ confidence levels.
  The results obtained using the galaxy, velocity, cross, and total data vectors are shown in blue, orange, cyan and red respectively.}
	\label{fig:cont_ideal}
\end{figure*}

Figure \ref{fig:cont_ideal} shows the one and two sigma confidence levels of the posterior distribution averaged over the 
25 realisations of the recovery set.
The true values of the parameters are indicated with the black lines.
The four methods give consistent results, both for the cosmological and HOD parameters, 
recovering every fitted parameters within one sigma.
The contours obtained using the different observables have slightly different shapes and inclinations. 
Except for the parameters that weakly affect the clustering and are therefore unconstrained, 
jointly fitting the galaxy and peculiar velocity statistics reduces parameter degeneracy and yields tighter constraints, 
especially for $w_0$ and $\sigma_8$.

\end{document}